\documentclass[fleqn,usenatbib]{mnras}

\usepackage{newtxtext,newtxmath}

\usepackage[T1]{fontenc}

\DeclareRobustCommand{\VAN}[3]{#2}
\let\VANthebibliography\thebibliography
\def\thebibliography{\DeclareRobustCommand{\VAN}[3]{##3}\VANthebibliography}

\usepackage{graphicx}	
\usepackage{amsmath}	
\usepackage{mathrsfs}
\usepackage[flushleft]{threeparttable}
\usepackage{siunitx}
\usepackage{booktabs, caption}
\usepackage{float}
\usepackage{pdflscape}
\usepackage{ulem}

\newcommand{\pyroa}{{\sc PyROA}}
\newcommand\mbh{M_{\bullet}}

\newcommand{\LCO}{{LCO}}
\newcommand{\pg}{PG~1119+120}
\newcommand{\msun}{\mathrm{M}_{\odot}}

\title[Echo Mapping of \pg]{Testing Super-Eddington Accretion onto a Supermassive Black Hole: Reverberation Mapping of \pg  }

\author[F. R. Donnan et al.]{Fergus R. Donnan,$^{1,2}$\thanks{E-mail: fergus.donnan@physics.ox.ac.uk}
Juan~V. Hern{\'a}ndez Santisteban,$^{1}$
Keith Horne,$^{1}$
Chen Hu,$^3$
Pu Du,$^{3}$
Yan-Rong Li,$^3$
\newauthor
Ming Xiao,$^3$
Luis C. Ho,$^{5,6}$
Jesús Aceituno$^{8,9}$
Jian-Min Wang,$^{3,4,6}$
Wei-Jian Guo$^{3}$
Sen Yang$^{3}$
Bo-Wei Jiang$^{3}$
\newauthor
Zhu-Heng Yao$^{3}$
\\
$^{1}$SUPA School of Physics and Astronomy, University of St~Andrews, North Haugh, St~Andrews KY16~9SS, Scotland, UK\\
$^{2}$Department of Physics, University of Oxford, Keble Road, Oxford, OX1 3RH, UK\\
$^{3}$Key Laboratory for Particle Astrophysics, Institute of High Energy Physics, Chinese Academy of Scienecs, 19B Yuquan Road, Beijing 100049, China \\
$^4$School of Astronomy and Space Sciences, University of Chinese Academy of Sciences, 19A Yuquan Road, Beijing 100049, Peoples Republic of China\\
$^5$Kavli Institute of Astronomy, Peking University, Beijing 100875, China\\
$^6$Department of Astronomy, School of Physics, Peking University, Beijing 100871, China\\
$^{7}$National Astronomical Observatories of China, Chinese Academy of Sciences, 20A Datun Road, Beijing 100020, Peoples Republic of China\\
$^8$Centro Astronomico Hispano Alemán, Sierra de los filabres sn, E-04550 Gergal, Almería, Spain\\
$^9$ Instituto de Astrofísica de Andalucía (CSIC), Glorieta de la astronomía sn, E-18008 Granada, Spain}

\date{Accepted XXX. Received YYY; in original form ZZZ}

\pubyear{2021}

\begin{document}
\label{firstpage}
\pagerange{\pageref{firstpage}--\pageref{lastpage}}
\maketitle

\begin{abstract}
We measure the black hole mass and investigate the accretion flow around the local ($z=0.0502$) quasar \pg. Spectroscopic monitoring with Calar Alto provides H$\beta$ lags and linewidths from which we estimate a black hole mass of $\log \left(\mbh/\msun \right) = 7.0$, uncertain by $\sim0.4$ dex. High cadence photometric monitoring over two years with the Las Cumbres Observatory provides lightcurves in 7 optical bands suitable for intensive continuum reverberation mapping. We identify variability on two timescales. Slower variations on a 100-day timescale exhibit excess flux and increased lag in the $u'$ band and are thus attributable to diffuse bound-free continuum emission from the broad line region. Faster variations that we attribute to accretion disc reprocessing lack a $u'$-band excess and have flux and delay spectra consistent with either $\tau \propto \lambda^{4/3}$, as expected for a temperature structure of $T(R) \propto R^{-3/4}$ for a thin accretion disc, or $\tau \propto \lambda^{2}$ expected for a slim disc. Decomposing the flux into variable (disc) and constant (host galaxy) components, we find the disc SED to be flatter than expected with $f_{\nu} \sim \rm{const}$. Modelling the SED predicts an Eddington ratio of $\lambda_{\rm Edd} > 1$, where the flat spectrum can be reproduced by a slim disc with little dust extinction or a thin disc which requires more dust extinction. While this accretion is super-Eddington, the geometry is still unclear, however a slim disc is expected due to the high radiation pressure at these accretion rates, and is entirely consistent with our observations. 
\end{abstract}




\section{Introduction}

Accretion onto supermassive black holes (SMBH) is thought to take place in a geometrically-thin and optically-thick accretion disc \citep[][]{Shakura1973} when accreting sufficiently below the Eddington limit. As the accretion rate approaches the Eddington limit, radiation pressure dominates, increasing the vertical thickness from a thin disc to a ``slim'' disc \citep[][]{Abramowicz1998}. The high accretion rate results in a strong radial motion, advecting photons onto the black hole before they can escape. Within this photon-trapping region, the scale height is significantly larger. Observational evidence in this regime is scarce, however it is thought to play a pivotal role in the early mass growth of SMBH's \citep[e.g. ][]{Du2016W,Pacucci2017, Regan2019}, potentially allowing black hole mass to build quickly in the early Universe. Understanding the processes behind such accretion is key to understanding black hole growth and the co-evolution of SMBHs and their host galaxies \citep[e.g.][]{Kormendy2013}.

As direct angular resolution of the accretion flow is unattainable with current technology,  indirect techniques such as echo (reverberation) mapping \citep[see][for a recent review]{Cackett2021} are required to probe the innermost regions of active galactic nuclei (AGN). This method trades angular resolution for temporal resolution, exploiting AGN variability and the finite travel time of light to dissect the accretion flow. Correlated variations observed with time delays between continuum lightcurves at different wavelengths are interpreted as evidence of thermal reprocessing in the surrounding region of X-ray/EUV irradiation originating very close to the black hole. Under this ``lamp-post'' model, longer-wavelength emission originates from larger radii and therefore appears later in time. The relationship of time delay with wavelength probes the temperature structure of the disc \citep[e.g ][]{Cackett2007}.

For the classical geometrically-thin disc \citep[][]{Shakura1973}, the temperature profile follows $T(R) \propto R^{-3/4}$ and the corresponding time delay increases with wavelength as $\tau(\lambda) \propto \lambda^{4/3}$, since the delay is $\tau \approx R/c$, where $c$ is the speed of light, and $T \propto 1/\lambda$ for blackbody emission.
A ``slim'' disc should show a different temperature profile with $T(R) \propto R^{-1/2}$ within the photon-trapping radius \citep[][]{Wang1999} and therefore a steeper delay spectrum, $\tau(\lambda) \propto \lambda^{2}$. These temperature profiles produce spectral energy distributions (SED) with very different power-law indices. A thin disc is expected to show $f_{\nu} \propto \lambda^{-1/3}$ whereas a slim disc should produce a redder SED, $f_{\nu} \propto \lambda^{+1}$. 

There are now many continuum reverberation mapping (RM) studies \citep[e.g. ][]{ Fausnaugh2016,Edelson2019,HernandezSantisteban2020} finding delay spectra compatible with $\tau\propto\lambda^{4/3}$ and thus consistent with the expected thin-disc temperature profile at sub-Eddington accretion rates. However, the inferred disc sizes are consistently larger than expected by typically a factor of 2-3 \citep[e.g.][]{Wei-Jian2022}. \citet{Cackett2020} performed the first intensive continuum RM study of for a super-Eddington quasar, and found a delay spectrum consistent with either a thin or slim disc but an SED that strongly followed $f_{\nu} \propto \lambda^{-1/3}$, expected for a thin disc temperature profile.

There are however numerous RM studies of the broad-line region (BLR) as part of the Super-Eddington Accreting Massive Black Holes (SEAMBH) collaboration, which were designed to understand physics of super-Eddington accretion and apply the saturated luminosity to cosmology \citep{Wang2013,Du2014,Du2019}. Measurements of the BLR size from H$\beta$ emission-line delays find sizes \citep[e.g.][]{Du2016, Du2018, FonsecaAlvarez2020} lower than expected compared to the traditional $R(L)$ relationship \citep{Kaspi2005, Bentz2013}, between the BLR radius ($R$) and the AGN luminosity at 5100\AA\ ($L$). This suggests a more compact BLR than for sub-Eddington AGNs with the same luminosity, and thus a different structure to the accretion flow. \citet{Wang2014} suggests that the geometrically-thick structure of slim accretion discs self-shadow the outer BLR to produce a highly anisotropic radiation field that with two distinct regions of the BLR, explaining the shortened lags. The highest accretion rates have been found to be $\sim 10^3\,\dot{M}c^2/L_{\rm Edd}$, where $\dot{M}$ is the mass accretion rate,
$c$ is the speed of light and $L_{\rm Edd}$ is the Eddington luminosity.

In this work, we perform continuum and emission-line reverberation mapping of the local $z=0.0502$ \citep{deVaucouleurs1991}, high-luminosity quasar \pg, first identified through the Palomar-Green survey \citep{Green1986}. Monitoring of this object was chosen as it is a potential super-Eddington accreting massive black hole \citep{Davis2011}, with an Eddington ratio $\sim 1$. In Section~\ref{sec:Obs} we describe the observations and data reduction. In Section~\ref{sec:TimeSeries} we describe how we model the lightcurves and determine inter-band delays. In Section~\ref{sec:Results} we present our results of the black hole mass and accretion disc analysis. In Section~\ref{sec:Discuss} we discuss the implication of our results for high luminosity AGN.

Throughout this work we assume $\Lambda$CDM cosmology with $H_0 = 67.8 \pm 0.9$ kms\textsuperscript{-1}Mpc\textsuperscript{-1} and $\Omega_m = 0.308 \pm 0.012 $ \citep[][]{Planck2016}.
The corresponding luminosity distance 
at the redshift $z=0.0502$ of \pg\ is $231$~Mpc.

\section{Observations}
\label{sec:Obs}
\subsection{Las Cumbres Observatory}

\subsubsection{7-band photometric monitoring}

The Las Cumbres Observatory (LCO) global robotic telescope network provided our 7-band photometric monitoring of \pg.  We achieved sub-day cadence over two observing seasons, 2019~Dec~1 through 2020~Jul~18 (Year~1) and 2020~Nov~1 through 2021~Jul~17 (Year~2).
\footnote{These observations were part of the LCO Key Projects KEY2018B-001 (PI. R. Edelson) and KEY2020B-006 (PI. J.~V. Hern\'andez Santisteban).} 
The Sinistro cameras on the
LCO 1-m robotic telescopes each cover a
$26.5\times26.5$~arcmin$^2$ field of view
with a $4K\times4K$~pixel array of $15\mu$-m CCD pixels
(0.389~arcsec~pixel$^{-1}$).
On each visit we took exposures in pairs to mitigate against cosmic ray impacts on the detector and to provide an internal consistency check on the error bar estimates.
Over both years, we achieved a median cadence of $\sim0.55$ days, with the Year~2 cadence closer to $\sim0.35$ days. 
The filter bands, typical exposure times, number of epochs, and median cadences are detailed in Table~\ref{tab:log_observations}.

We downloaded CCD image data from the LCO archive\footnote{\url{https://archive.lco.global/}} which provides bias and flat-field corrected images automatically processed by the {\sc banzai} \citep{curtis_mccully_2018_1257560} pipeline. 
We extracted multi-aperture photometry with {\sc SExtractor} \citep{bertin:1996} on every image. We constructed a global background model by smoothing the image in a 200 pixel mesh, large enough to avoid the extended sources influencing the background estimate. After subtracting the background model, we performed aperture photometry with a $5\arcsec$ radius aperture, large enough to produce robust light curves against a range of atmospheric conditions (e.g., airmass, seeing) taken throughout the year and in different sites but small enough as to not compromise signal-to-noise ratio. We used comparison stars in each field to perform an image zero-point calibration at each epoch. We used the \textit{AAVSO Photometric All-Sky Survey} (APASS) DR10 \citep{henden:2018} for \textit{g$^\prime$, r$^\prime$, i$^\prime$}, $B$ and $V$ filters, and for  \textit{z$_s$} we used Pan-STARRS1 \citep{Flewelling2020}. For the \textit{u$^\prime$}~band, where no APASS information was available, we made use of the \textit{Swift}/UVOT \textit{U} band images to obtain the fluxes of reference stars in the field. All colour-correction and atmospheric extinction coefficients were obtained from \citet{valenti:2016} and applied before the photometric calibration. We applied a 3-$\sigma$ clipping to the zero-point estimates and used bootstrap samples to estimate uncertainties.
\begin{table}
	\centering
	\caption{Photometric and spectroscopic monitoring of \pg\ with \LCO\ and Calar Alto.}
	\label{tab:log_observations}
    \def\arraystretch{1.2}
    \setlength{\tabcolsep}{4pt}
	\begin{tabular}{cccccc}
	
	\multicolumn{5}{c}{{\bf LCO Photometry}}\\
	\hline
	1-m/Sinistro & $\lambda_{\rm eff}$ & FWHM & $t_{\rm exp}$ & Epochs & Cadence\\
	Filter & (\AA) & (\AA) & (s) &  & (days)\\
	\hline
	{$u^\prime$}   & 3580 & 570 & $2\times300$ & 398 & 0.51  \\
	{$B$}            & 4392 & 890 & $2\times60$ & 864 & 0.52  \\
	{$g^\prime$}   & 4770 & 1500 & $2\times60$ & 1064 & 0.60  \\
	{$V$}            & 5468 & 840 & $2\times60$ & 856 & 0.62  \\
	{$r^\prime$}   & 6215 & 1390 & $2\times60$ & 953 & 0.64  \\
	{$i^\prime$}   & 7545 & 1290 & $2\times60$ & 1008 & 0.80  \\
	{z$_s$}        & 8700 & 1040 & $2\times120$ & 909 & 0.66  \\
	\hline\\
	
	\multicolumn{5}{c}{{\bf LCO and Calar Alto Spectroscopy}}\\
	\hline
	Spectrograph & $\lambda$ & $\Delta\lambda$ & $t_{\rm exp}$ & Epochs & Cadence\\
	 &   (\AA) & (\AA~pix$^{-1}$) & (s) &  & (days)\\
	 \hline
	FLOYDS-Red & 5400-10000 & 3.51 & $2\times1200/900$ & 32 & 3.2 \\
	FLOYDS-Blue & 3200-5700 & 1.74  & $2\times1200/900$ & 32 & 3.2 \\
	CAFOS-G200  & 4000-8000 & 4.47 & $2\times600$ & 32 & 6 \\
	\hline
	\end{tabular}
\end{table}

\subsubsection{Inter-telescope calibration}

With multiple telescopes contributing data, the observations from different telescopes must be intercalibrated and merged to form a single lightcurve for a given filter. The inter-telescope calibration determines telescope-specific scale factors $A_s$, which can arise from differences in the CCD sensitivities and the filter and telescope transmissions, and additive background flux offsets $B_s$, which may arise from different angular aperture sizes or angular resolution. To optimise the inter-calibration parameters we use \pyroa\footnote{\url{https://github.com/FergusDonnan/PyROA}} \citep{Donnan2021}, which models the underlying lightcurve, $X(t)$, as a running optimal average (ROA) that is scaled by a factor $A_s$ and shifted by $B_s$ to fit the flux data over all epochs for telescope $s$:
\begin{equation}
\label{eqn:InterCalibModel}
    f_{s}(t) = A_{s}\,X\left(t \right) + B_{s}
    \ .
\end{equation}
Here $X(t)$ and $B_s$ carry the same units as the lightcurve flux data and $A_s$ are dimensionless scale factors.
By optimising the scale and shift parameters for each telescope, the telescope-specific flux offsets are then removed to produce a merged lightcurve. \pyroa\ uses MCMC sampling of the lightcurve shape and calibration parameters. $X(t)$ is calculated as the ROA of all the individual lightcurves shifted and merged, thus tapping all available information on the lightcurve shape. The timescale parameter, $\Delta$, that controls the flexibility of the ROA model is optimised by computing the corresponding effective number of parameters and minimising the resulting Bayesian Information Criterion (BIC). 

 To compensate for some of the nominal uncertainty estimates being too small, the \pyroa\ noise model includes telescope-specific extra variance parameters, $\sigma_s^2$, which are added in quadrature with the nominal flux uncertainties. While
 $\chi^2$ then decreases with $\sigma_s$, the log-likelihood penalises models with expanded error bars, thus enabling $\sigma_s$ to be determined for each telescope. We use a uniform prior for $\sigma_s$. 
 
 As the LCO 1-m telescopes and Sinistro cameras are nearly identical, the the scale factors $A_s$ should be close to 1 and the background flux shifts $B_s$ close to 0.
 We adopt appropriate priors to implement these these expectations.
 For $A_s$ we adopt a narrow log-Gaussian prior:
\begin{equation}
    \textrm{Pr}(\ln A_s) = \frac{1}{0.02 \sqrt{2 \pi }} \exp\left[-\frac{1}{2} \left (\frac{\ln A_s}{0.02} \right)^2\right]
    \ .
\end{equation}
This log-Gaussian prior ensures that the $A_s$ remain positive. The fractional width 0.02, corresponding to a rms difference of 2\% between the lightcurves, is chosen based on the lightcurves of calibration stars that are not intrinsically variable. For the background flux shifts $B_s$ we adopt a Gaussian prior with a mean of 0 and a width of 0.5 mJy:
\begin{equation}
    \textrm{Pr}( B_s) = \frac{1}{\sqrt{2 \pi } \,0.5 \,\textrm{mJy} } \exp\left[-\frac{1}{2} \left (\frac{B_s}{0.5 \,\textrm{mJy}} \right)^2\right]
    \ .
\end{equation}

The \pyroa\ fit also implements a ``soft'' sigma clipping, whereby the error bars of outliers are expanded by a factor to keep their residuals no larger than $4\sigma$. Expanding the error bars, rather than simply removing the outlier data, prevents a discontinuity in $\chi^2$ by keeping outlier contributions to $\chi^2$ constant beyond $4\sigma$ rather than dropping abruptly to zero.

\begin{figure*}
\hspace*{-0.5cm}                                                           
    \centering
	\includegraphics[width=18cm]{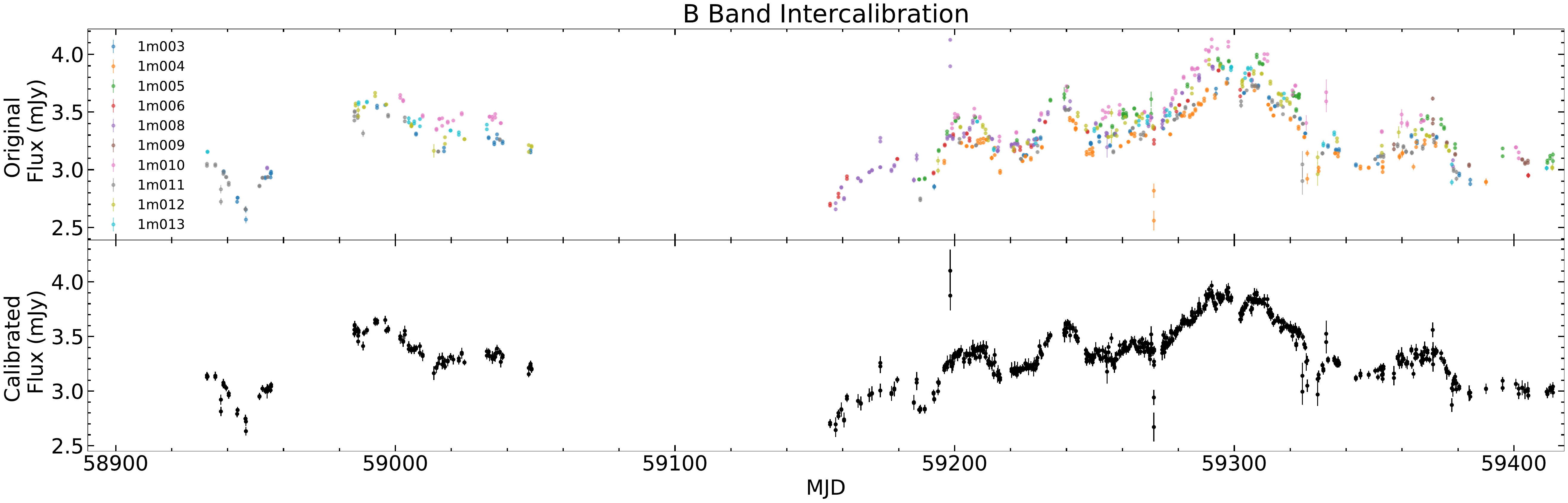}
    \caption{LCO B band observations for \pg, shown before (top panel) and after (bottom panel) intercalibration. In the top panel, each colour represents a single telescope from the Las Cumbres Observatory where they have been merged into a single lightcurve in the bottom panel. Note the expanded error bars from the noise model and the outliers from the sigma clipping.}
    \label{fig:InterCalib}
\end{figure*}

 Fig.~\ref{fig:InterCalib} presents an example of the \pyroa\ inter-telescope calibration, where the scatter present in the $B$-band lightcurve combining data from 10 LCO 1-m telescopes is reduced after making the telescope-specific corrections.

\subsubsection{FLOYDS spectroscopic monitoring}
In parallel with our 7-band photometric monitoring with the LCO 1-m network,
we deployed the FLOYDS spectrographs on the LCO 2-m robotic telescopes
(FTN at Haleakala, Hawaii and FTS at Siding Springs, Australia) for spectroscopic monitoring. 
FLOYDS\footnote{\url{https://lco.global/observatory/instruments/floyds/}} 
is a cross-dispersed low-resolution spectrograph covering the 3200 to 10,000~\AA\ spectral range by imaging 1st-order blue and 2nd-order red spectra on the same CCD.
The H$\beta$ region is imaged in both orders.
The 30-arcsec long slit, oriented to the parallactic angle to minimise the impact of differential refraction, provided for subtraction of the sky spectrum.
In total, we obtained 32 spectra, taken in pairs, with 1200-s exposures and a $2\arcsec$ slit width in Year~1 and 900-s exposures and a $6\arcsec$ slit width in Year~2. Wavelength and flat-field lamp calibrations are taken before every exposure. We employed the {\sc AGNFLOYDS} pipeline\footnote{\url{ https://github.com/svalenti/FLOYDS pipeline}} to reduce the data. To further improve flux calibration, remove telluric lines and minimise fringing artefacts (which are significant at wavelengths larger than 6000\AA), we used the closest standard star observed within 5 days.

\subsection{Calar Alto}

We obtained $V$-band images and long-slit spectra with the 2.2-m telescope at
Calar Alto Observatory for 32 epochs between 2020~Nov and 2021~Jul. The observing
strategy is the same as described in
\citet{Hu2021}. For each epoch, three broad-band photometric images with a 
Johnson $V$ filter and then two spectra with Grism G-200 and a long slit set
at a width of 3$\farcs$0 were taken by the Calar Alto Faint Object
Spectrograph (CAFOS). A nearby star (non-varying, confirmed by our photometric
observations) was taken simultaneously with \pg\ for flux calibration, by
rotating the slit. The dispersion and the exposure time of the spectra, along
with the number of epochs and the cadence, are listed in Table 
\ref{tab:log_observations}.  More details of the observations, data reduction,
and flux calibration can be found in \citet{Hu2021}.

For each calibrated spectrum, the 5100 \AA\ continuum flux was measured in the
wavelength window of 5085--5115 \AA\ (in the rest frame, so as the two wavelength windows mentioned below). A local continuum was defined as a straight
line by the 5100 \AA\ continuum window and another window of 4750--4780 \AA.
Then the H$\beta$ flux was measured as the integration over the range of
4810--4910 \AA\ above this continuum.

\subsection{Swift}

We obtained observations with the \textit{Neil Gehrels Swift Observatory} \citep[\textit{Swift} hereafter,][]{Gehrels04} in two visits: 
2021~Jun~26 and Jul~5, for a combined exposure time of 1.923~ks. We used the 0x224c mode in order to obtain measurements in all 6 UV/optical filters. We extracted the photometry using {\sc uvotsource} with an aperture of $5\arcsec$ radius centred on the target and a $30\arcsec$ background offset region, close to the AGN.  

We combined the two visits to create an average X-ray spectrum with XRT, extracted with the Swift online tool  \citep{Evans2009} using {\sc heasoft} v6.28. The combined spectrum resulted in a total of 357 detected photons in the full 1.83~ks exposure (no photons above 7~keV were detected above the background), for an average count rate of 0.18~counts~s$^{-1}$.

\section{Time series analysis methods}
\label{sec:TimeSeries}

The primary aim of our time-series analysis is to estimate inter-band time delays, as well as the mean and rms flux at each wavelength. This evidence, from both the lag spectrum, $\tau(\lambda)$, and the spectral energy distribution, $f_\nu(\lambda)$, of the variations, then enables tests of the thin-disc and slim-disc models that predict distinct disc temperature profiles, $T(r)\propto r^{-3/4}$ and $\propto r^{-1/2}$, respectively.
We describe the methods below and present the results in Section.~\ref{sec:Results}.

\subsection{\pyroa}

We use \pyroa\ \citep[][]{Donnan2021} to model the flux data of the 7-band LCO lightcurves and determine the inter-band delays. Analysing each year of data separately, we measure delays relative to the $g'$-band lightcurve, which has more epochs and higher signal-to-noise ratio than the others. \pyroa\ models the lightcurve shape with a running optimal average, and uses MCMC sampling to obtain accurate uncertainty estimates on the lightcurve parameters. This method models all of the lightcurves simultaneously using all of the data available to determine the shape of the variability, providing an advantage over cross-correlation methods that estimate lags from pairs of lightcurves. Full details of the fitting process can be found in \citet{Donnan2021}. 

We use \pyroa\ to model the lightcurves in 3 different ways increasing in complexity. In each case, \pyroa's running optimal average provides a dimensionless lightcurve, $X(t)$, that is normalised to zero mean and unit rms: $\left< X\right>_t=0, \left< X^2 \right>_t=1$. The ROA parameter, $\Delta$, gives the width of the Gaussian window function and controls the flexibility of the lightcurve model.

In Model I, the simplest, $X(t)$ is shifted to a mean flux $B_i$, scaled to an rms flux $A_i$, and translated in time by the delay $\tau_i$.
The Model I flux of lightcurve $i$ is then
\begin{equation}
\label{eqn:model1}
   \textrm{Model~I:} \quad f_{i}(t) = A_{i}\,X\left(t - \tau_{i}\right) + B_{i} \ ,
\end{equation}
where $A_{i}$ represents the rms flux, $B_{i}$ represents the mean flux, $\tau_{i} $ represents the time delay, and $X(t)$ is the driving lightcurve. The time delay for the $g'$ band lightcurve is fixed at zero therefore all the delays are measured relative to this band. This is known as Model~I hereafter. 

One notable feature of AGN lightcurves are variations on different timescales. While the fast variations are typically reprocessing of the driving X-ray/EUV flux, slower variations may be from some other physical process, which if unaccounted for, may lead to unreliable results. In particular \citet{HernandezSantisteban2020} and \citet{Vincentelli2021} observed slow variations behaving differently to the fast variations, which reverberated as predicted by the ``lamp-post'' model. Year 2 of the lightcurves for \pg\ shows a slow rise and fall in flux over the observing window - something unable to be seen in year 1. Therefore for the Year 2 data we extend Model~I to include a slow varying component for each lightcurve, $S_i(t)$, which is given by a parabola of the form
\begin{equation}
\label{eqn:SlowComp}
    S_i(t) = f_0 + \Delta f \left( \frac{t - t_0}{100 \textrm{days}}\right)^2 \ ,
\end{equation}
where $f_0$ is the peak flux at time $t_0$ and $\Delta f$ represents the change in flux over 100 days. This is fitted to each lightcurve, and then normalised to have a mean of zero over the range of data, before it is subtracted from each lightcurve during the fitting procedure of \pyroa, leaving only the fast variations to be modelled and used to determine the time delays. This model is given by 
\begin{equation}
\label{eqn:model2}
   \textrm{Model~II:} \quad f_{i}(t) = A_{i}X\left(t - \tau_{i}\right) + B_{i} + S_i(t) \ ,
\end{equation}
and is known as Model~II hereafter. 

Both Models I and II, assume that the shape of the variability is the same and simply shifted by a single time delay between lightcurves. In reality, the emission of an AGN accretion disc at a given wavelength will originate from various different radii, with the peak flux originating at a certain wavelength given by the temperature structure of the disc. The means the lightcurve at a given wavelength is constructed from a distribution of delays that is convolved with the driving lightcurve. Under this thermal reprocessing model, the exact shape of the delay distributions as a function of wavelength, depend on various disc properties such as the temperature structure, the mass/accretion rate $M\dot{M}$, and the inclination $i$ \citep[e.g][ and Section~\ref{sec:cream} for detailed accretion disc modelling]{Starkey2016}. We therefore extend Model~II by including a convolution with a delay distribution, $\Psi_i(\tau)$, giving
\begin{equation}
\label{eqn:model3}
   \textrm{Model~III:} \quad f_{i}(t) = A_{i} \int_{-\infty}^{\infty} \Psi_i(\tau) X\left(t - \tau_{i}\right) \,d\tau + B_{i} +S_i(t) \ ,
\end{equation}

where $\tau_i$ now represents the mean delay. This has the effect of smoothing out the variations where the wider the delay distribution, the smoother the lightcurve. This simplifies to Eqn.~(\ref{eqn:model2}), where the delay distribution is a Dirac delta function, $\delta(\tau)$, which specifies a single time delay. While models I and II may be adequate to obtain time delays, modelling the delay distribution may provide a better fit to the data. We include the slow varying component when extending the model further, motivated by the better BIC of Model~II vs Model~I (see Section~\ref{sec:DelaySpec}) as well as the physical justification for the component in Section~\ref{sec:diffuseBLR}.

To implement this process into \pyroa\ we calculate a new window function from the convolution of the original Gaussian window function, $w_i(t)$, with the delay distribution, $\Psi_i(\tau)$, for each lightcurve, $i$. As the running optimal average is calculated from the data and is not a prior on the shape of the driving lightcurve, unlike a damped random walk for example, the delay distributions must be measured relative to the shortest wavelength band. Therefore the ROA for the reference band, in this case the $u'$ band, is calculated using the original Gaussian window function and the remaining window functions are this Gaussian convolved with some delay distribution, $\Psi_i(\tau)$. 

There are many choices one could make for the form of $\Psi_i(\tau)$, but in this case we use a log-Gaussian of the form
\begin{equation}
\label{eqn:DelayDist}
    \Psi_i(\tau) = \frac{1}{\tau \sigma_i} \exp \left[ - \frac{\left( \ln \tau - \mu_i \right)^2}{2 \sigma_i^2}\right], 
\end{equation}
where the parameters $\mu_i$ and $\sigma_i$ are given in terms of the mean delay, $\tau_i$, and the rms around this mean, $\tau_{\rm{rms}, i}$
\begin{equation}
\label{eqn:DelayDistConv}
    \mu_i = \ln \left( \frac{\tau_i^2}{\sqrt{\tau_i^2 + \tau_{\rm{rms}, i}^2}}\right), \quad \sigma_i^2 = \ln \left( 1 + \frac{ \tau_{\rm{rms}, i}^2}{\tau_i^2}\right) .
\end{equation}
The mean delay, $\tau_i$ is sampled as before, as well an additional parameter per lightcurve - the rms of the log-Gaussian delay distribution, $\tau_{\rm{rms}, i}$. This allows the mean and width of~$\Psi_i(\tau)$ to vary freely for each lightcurve. The use of a log-Gaussian prevents delays less than zero from contributing which enforces causality. As the minimum delay allowed within this log-Gaussian distribution is simply the sampled $\tau_i$ for $i=0$, this may not necessarily be zero if $i=0$ is not the reference for the time delay. Indeed we measure the mean delays relative to the $g$ band in this work while the blurring is measured relative to the $u'$ band. Therefore when calculated in the code, we calculate the log-Gaussian with a mean measured relative to the minimum delay, interpolate and shift back such that the ``zero'' point of the log-Gaussian is at the minimum delay.

We fit each model to each year of data separately, using only Model~I for the first year and additionally using models II and III for the second year, which we focus on in this paper. For models I and II we used 25~000 samples, 58 walkers and discarding the initial 20~000 as burn-in, and 76 walkers for Model~III. A full list of the best fit parameters can be found in Table. \ref{tab:PyROAParameters}.

\subsection{Cross-correlation function}
\label{sec:CCF}
Finally we used the interpolation cross-correlation function (ICCF) \citep{ICCF} to further measure the inter-band delays using the code {\sc PyCCF} \footnote{\url{https://bitbucket.org/cgrier/python_ccf_code/src/master/}} \citep{PyCCF}. To avoid the large gap in the data, we measured delays for each year separately. As our lightcurves featured outliers that had expanded error bars from the sigma clipping during the intercalibration stage, the ICCF had difficulty managing these outliers as this method relies on interpolating the lightcurve which meant outliers create distinct features that may distort the lag measurement. Therefore these outliers were removed by taking the model ROA from the intercalibration stage and removing any data points outwith $3\Bar{\sigma}$, where $\Bar{\sigma}$ is the mean error on the flux. We used the mean error rather than each individual error in order to remove outliers that had large error bars which would not be outwith $3\sigma$ but would still cause extraneous features in the interpolated lightcurve. 

To allow comparison with \pyroa, we again used the $g'$ band as the reference lightcurve. We used an interpolation grid between -15 and 15 delays, which specifies the allowed lag range and a spacing of 0.1 days. The errors for this method are estimated using flux randomisation/random subset selection (FR/RSS) \citep{Peterson1998} where delays are measured from multiple realisations of the CCF. We used the centroid distributions of the CCF where values of the CCF $> 0.8\,r_{\textrm{max}}$ were used where $r_{\textrm{max}}$ is the maximum value of the CCF. As the $u'$ band lightcurve does not fully overlap with the $g'$ band, with a gap at the start, only the overlapping regions were used when measuring the $u'$ band delay.

To compare with the \pyroa\ results we determined delays using the ICCF for the original data as well as accounting for a slow varying component. This was done by fitting Eqn.~(\ref{eqn:SlowComp}) to each lightcurve, which is then normalised to have a mean of zero over the range of data and is then subtracted from each lightcurve, leaving only the fast variations. We then use the PyCCF code on the resulting lightcurves, measuring delays only using the fast variations. 

To measure the black hole mass in section \ref{sec:BHMass}, we similarly use the ICCF method to measure time delays of H$\beta$ relative to the continuum. Here we do not subtract any slow component from the lightcurves before inferring the delays.

\section{Results}
\label{sec:Results}
\subsection{Black Hole Mass}
\label{sec:BHMass}

\begin{figure*}
    \centering
    \includegraphics[width=2\columnwidth]{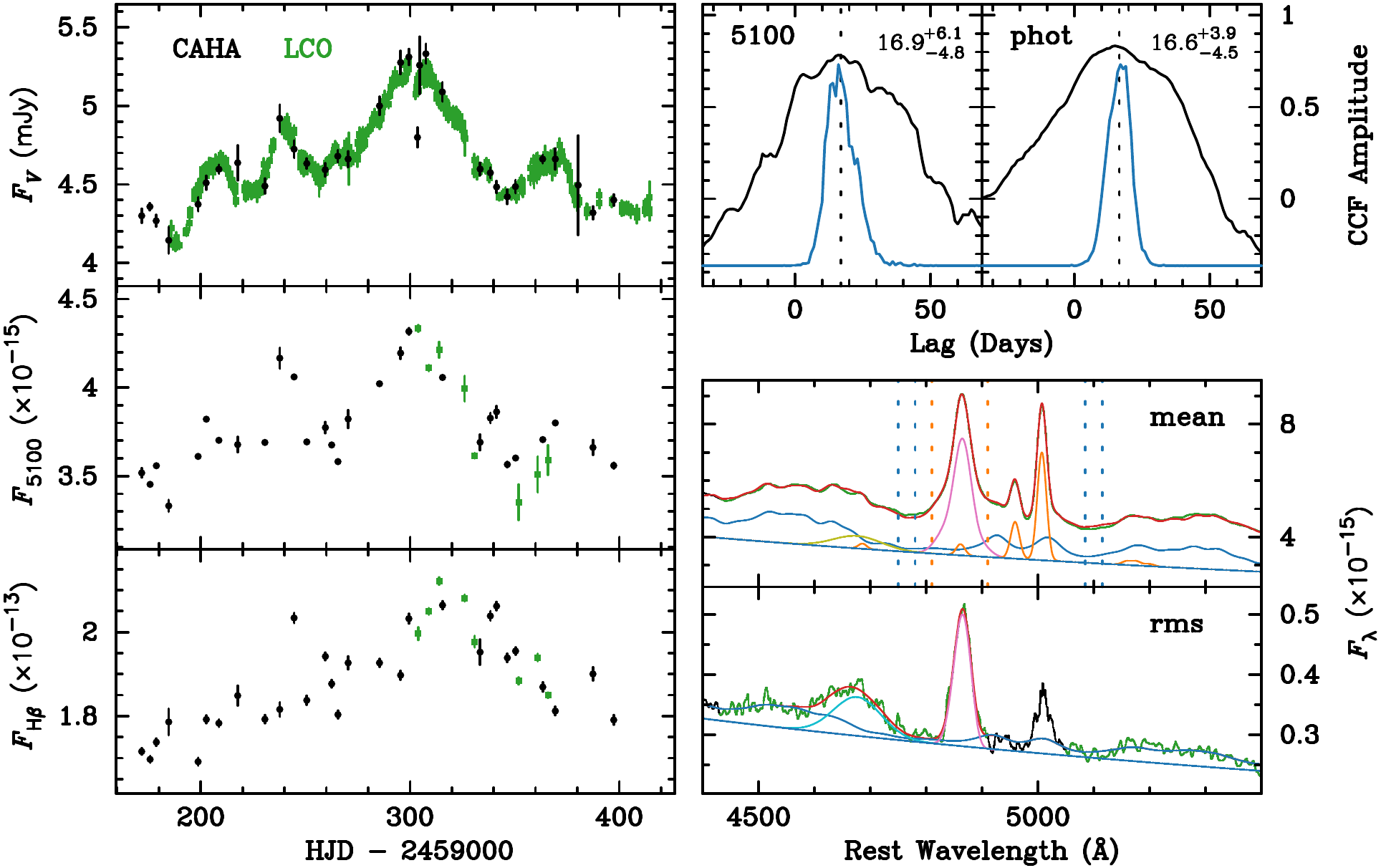}
    \caption{H$\beta$ emission-line lag measurements based on the CAHA data.
    \textit{Left column:} lightcurves from $V$-band photometry (top), 5100\AA\ continuum spectroscopy (middle) and continuum-subtracted H$\beta$ emission-line flux (bottom). Both $F_{5100}$ and $F_{{\rm H}\beta}$ are in units of erg s$^{-1}$ cm$^{-2}$ \AA$^{-1}$. \textit{Right column top:} Cross-correlation functions (black) and centroid lag distributions (blue). The inferred rest-frame time lags are shown in the top right corners of the CCF plots. \textit{Right column bottom:} The Mean and RMS spectra and the best-fit model including a power-law continuum, Fe~II template, and Gaussian emission-line profiles for He~II, H$\beta$ and [OIII]. Both mean and RMS spectra are in units of erg s$^{-1}$ cm$^{-2}$ \AA$^{-1}$.}
    \label{fig:hblag}
\end{figure*}

In Fig.\ref{fig:hblag} we show the main results of our
CCF analysis of the 2021 CAHA data.
Spanning 190 days,
the $V$-band photometric and 5100~\AA\
continuum lightcurves exhibit a similar pattern
of variations, with several
minima and maxima on a rising trend.
The strongest feature suitable for time-delay measurements is the maximum near
$t={\rm JD}-2459000=300$~d and subsequent fall by 20\% to a minimum near $t=350$~d.
The H$\beta$ lightcurve records fluctuations
on a rising trend, peaking near $t=340$~d, 
followed by a drop of 10\% by $t=370$~d when the
lightcurve ends.
Inspecting these lightcurves, 
the relative timing of the final maximum and subsequent decline suggests a 
lag of $20-30$~d for H$\beta$ relative to the continuum variations.

The cross-correlation analysis presented in
the top-right panel of Fig.\ref{fig:hblag}
provides the results summarised in Table~\ref{tab:lags}. The V band is supplemented with the much higher cadence LCO photometric observations. For the LCO spectroscopy, a number of epochs did not coincide with the CAHA observations and some of the spectra showed fringing effects around H$\beta$ leading to only 8 additional epochs from the LCO data to the spectroscopic lightcurves shown in the middle and bottom panels of Fig.~\ref{fig:hblag}.

The peak correlation coefficients $r_{\rm max}>0.7$ are high enough to confirm the eyeball impression of correlated variations. The CCF peaks are shifted,
indicating that the H$\beta$ variations are delayed,
by about $19\pm7$~d
relative to the $V$-band lightcurve,
and $19\pm14$~d relative to the 5100~\AA\ continuum
lightcurve.

\begingroup
\setlength{\tabcolsep}{10pt} 
\renewcommand{\arraystretch}{1.5} 
\begin{table}
    \centering
    \begin{tabular}{cccc}
    \hline
    Lightcurve & Line & $r_{\rm max}$ & $\tau$  \\ 
     & & & (Days)\\\hline
       5100\AA\  &  H$\beta$ & 0.78 & $16.9_{-4.8}^{+6.1}$\\
        $V$-band &H$\beta$ & 0.83 & $16.6_{-4.5}^{+3.9}$ \\ \hline
    \end{tabular}
    \caption{H$\beta$ cross-correlation peak $r_{\rm max}$ and rest-frame time lag $\tau$ (days) from the CCF analysis presented in Fig.~\ref{fig:hblag}.}
    \label{tab:lags}
\end{table}
\endgroup

Fits to the Mean and RMS spectra, using a multi-component model including a power-law continuum, Fe~II template spectrum convolved with a Gaussian velocity profile,
and Gaussian velocity profiles for He~II, H$\beta$ and the two [OIII] emission-lines, are shown in the lower-right panel of Fig.~\ref{fig:hblag}.
The fit is good, and provides H$\beta$ linewidth estimates, both FWHM and RMS, as summarised 
in Table~\ref{tab:BHMass}. These linewidths are corrected for the instrumental resolution. We only use the CAHA spectra to measure the H$\beta$ linewidth as including the LCO spectra may introduce additional uncertainty due to the different spectral resolution and wavelength range. 

The RMS spectrum shows some residual [OIII] features indicating imperfect spectrophotometric corrections.
The residual Fe~II features in the RMS spectra may similarly be a calibration artefact.



Based on measurements of the H$\beta$ lag $\tau$, 
the H$\beta$ radius is $R\sim\tau c$, and
the black hole mass is then
\begin{equation}
    \mbh = f \,\frac{\Delta V^2 \ \tau c}{ G } \ .
\end{equation}
The dimensionless factor $f$ accounts for systematic error due to our ignorance of the geometry, kinematics, and orientation of the BLR gas in each AGN.
Calibrated estimates for the population-mean are
is $\langle f\rangle\sim5$ if $\Delta V$ is the RMS linewidth \citep[e.g.][]{Grier2017}, and $\langle f\rangle\sim1$ if $\Delta V$ is FWHM \citep[e.g.][]{Woo2015}. The uncertainty in $f$ for individual AGN is circa 0.4~dex, and this dominates the error budget, given our roughly 10\% uncertainty in $\Delta V$ and 30\% uncertainty in $\tau$.

\begin{table}
\centering
  \caption{Virial products for CAHA line widths and a H$\beta$ lag of $16.6_{-4.5}^{+3.9}$ days.}
  \label{tab:BHMass}
    \def\arraystretch{1.2}
    \setlength{\tabcolsep}{6pt}
    \begin{threeparttable}
  \begin{tabular}{cccc}
  
    \hline
    Spectrum & Linewidth (Type) & $\mbh/f$ & $\log \left(\mbh/\msun \right)^{\dagger}$  \\
    & (km~s$^{-1}$) & ($10^7 \msun$) & \\

    \hline
    Mean & $2244\pm174$ (FWHM) & $1.61 \pm 0.48$ & 7.21 \\
    Mean & $1317\pm37$ (RMS) & $0.56 \pm 0.14$&  7.45 \\
    RMS & $1911\pm165$ (FWHM) & $1.16 \pm 0.36$ & 7.01\\
    RMS & $768\pm88$ (RMS) & $0.187 \pm 0.065$& 6.97 \\
   \hline
  
  \end{tabular}
\begin{tablenotes}
    \item[$\dagger$] Calculated using $f=5$ for RMS linewidth and $f=1$ for FWHM. The uncertainty in $\log \left(\mbh/ \msun\right)$ is $\sim 0.4$ as a result.
  \end{tablenotes}
  \end{threeparttable}
 \end{table}

Table~\ref{tab:BHMass} shows the virial product $\mbh/f$, for the linewidths from the CAHA spectra and using a H$\beta$ lag of $16.6_{-4.5}^{+3.9}$ days. We used the delay relative to the $V$ band as this was better constrained, utilising the high cadence LCO observations. Assuming values of $f$ described previously, we find a range of $\log \left(\mbh/\msun \right)$ between 6.97 and 7.45, all with uncertainty of $\sim 0.4$ dex. As the rms spectrum better isolates emission from the AGN, the lower two mass estimates are likely more accurate. For the subsequent analysis in this paper, we use the measured mass from the FWHM linewidth for the rms spectrum, giving a mass of $\log \left(\mbh/\msun \right)= 7.0$.


\subsection{X-ray Spectroscopy}
\label{sec:XSpec}
\begin{figure}
	\includegraphics[trim=0cm .1cm 0.2cm 0cm,clip,width=1\columnwidth]{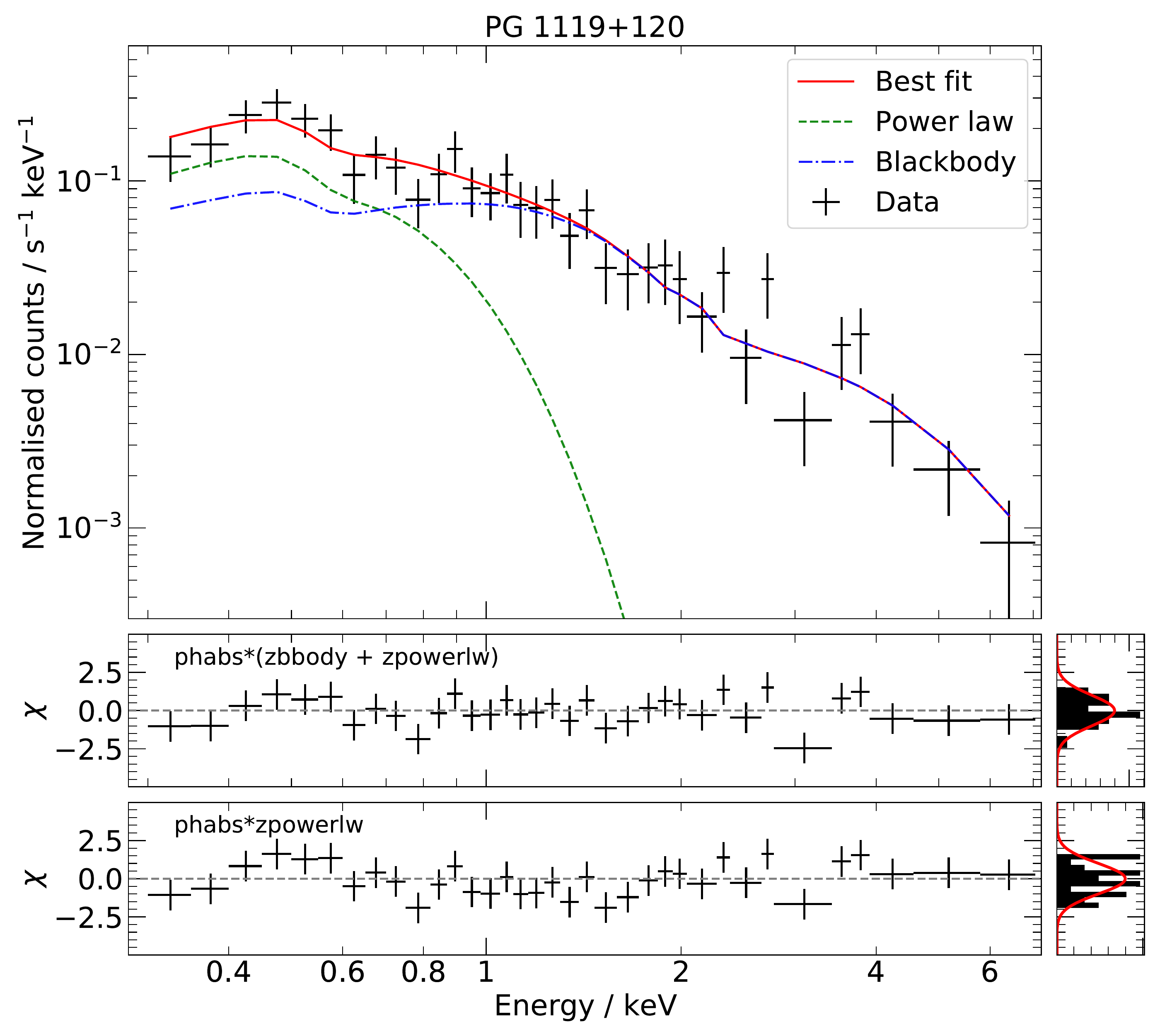}
    \caption{Average X-ray spectrum of \pg\ and best fit model during the 2021 campaign. {\it Top:} The individual model components of the best fit (red) are a blackbody (green) and power-law (blue). The residuals for the two models performed are shown in the middle and bottom panels. In the right-hand panels, a histogram of the normalised residuals are shown in comparison to a standard Gaussian distribution.}
    \label{fig:xray_spectrum} 
\end{figure}

We made an initial fit with {\sc Xspec} \citep{Arnaud1996} to the 0.3-7 keV unbinned spectrum with a power-law model attenuated by Galactic absorption -- {\sc phabs*zpowerlw}. Due to the low number of photons, we used C-stat \citep{Cash1979,Kaastra2017} as our goodness-of-fit statistic. The uncertainties in the individual parameters (68\% confidence interval) were calculated by a MCMC procedure with $10 000$ iterations and 15 walkers within {\sc Xspec}.
We used the line-of-sight Galactic optical extinction E(B-V)=0.033 \citep{Schlafly2011} to fix the total hydrogen column density in this analysis to N$_{H} = 2.26\times10^{20}$ cm$^{-2}$, using the \citet{Guver2009} relations. 
The power-law fit resulted in a good description of the overall broadband spectrum with a C-stat/degree of freedom (d.o.f.) = 148.28/170. However, at lower energies, clear modulations are observed in the residuals (middle panel of Fig.~\ref{fig:xray_spectrum}). We added an additional blackbody component to better constrain the low-energy range, thus the model in {\sc Xspec} format is {\sc phabs*(zbbody+zpowerlw)}. This resulted in an improved statistical fit, with a C-stat/d.o.f$= 137.80/168$, with a change of $\Delta$C-stat$=-10.48$ for 2 additional d.o.f, shown in Fig.~\ref{fig:xray_spectrum}. The best fit results for the two models considered are found in Table~\ref{tab:xrays} as well as the extrapolated flux in the 2-10 keV band, $F_{X,2-10} = 2.824\times10^{-12}$ erg s$^{-1}$ cm$^{-2}$. Thus, the luminosity of in this band is $L_{X,2-10} = 1.67\times10^{43}$ erg s$^{-1}$. 
Using, the bolometric correction ($k_{\rm BOL}=53$) from \citet{Netzer2019}, we find a bolometric luminosity of $L_{\rm BOL} = 9.00\times10^{44}$ erg s$^{-1}$. This suggests that \pg\ has an Eddington luminosity ratio (with a $\log \left(\mbh/\msun \right) = 7.0$) of $\lambda_{\rm Edd} = 0.71$. This estimate is lower than the analysis from the optical SED measurements (see Section~\ref{sec:accretionrate}.
Given the low signal-to-noise of our spectrum, we didn't pursue to fit more physically motivated models, as our main goal was to retrieve the flux in the 2-10 keV band for the delay spectrum analysis in Section~\ref{sec:accretionrate}. 



\begin{table*}
  \caption{Best fit parameters for the average X-ray spectrum of \pg. Fixed parameters are labelled with $^{*}$. Both models have been multiplied by {\sc phabs} using the N$_{\rm H}$ as a fixed parameter. }
  \begin{centering}
  \begin{tabular}{lcccc}
  \hline
Parameter & units& {\sc zpowerlw}&{\sc zbbody+zpowerlw} & Description\\
    \hline
N$_{\rm H}$ & $\times10^{20}$ cm$^{-2}$&2.26$^{*}$ & 2.26$^{*}$& Column Density\\
$\Gamma$ && $2.43\pm0.08$ & $1.95\pm0.21$ & Photon index\\
$N_{PL}$ & $10^{-3}$& $1.62\pm 0.09$ & $1.07\pm0.24$ & Power law normalisation\\
kT &keV& $\dotsm$ &$0.11\pm0.02$ & Blackbody Temperature \\
$N_{BB}$ & $10^{-5}$& $\dotsm$ & $4.2\pm1.2$ & Blackbody normalisation \\
C-stat/d.o.f. &  & $148.28/170$  & $137.80/168$ &  Fit Statistic\\
\hline
$F_{X,0.3-2}$ & erg s$^{-1}$ cm$^{-2}$ & 4.2622e-12 & 4.1859e-12 & Observed flux in the $0.3-2$ keV band\\
$F_{X,2-10}$ & erg s$^{-1}$ cm$^{-2}$ & 1.9554e-12 & 2.8254e-12 & Observed flux in the $2-10$ keV band\\
\hline
  \end{tabular}

  \label{tab:xrays}
  \end{centering}
\end{table*}

\subsection{Delay Spectrum}
\label{sec:DelaySpec}
The delays measured by \pyroa\ and the CCF, for each year of data are shown in Table~\ref{tab:Results}. The CCF results corresponding to Model~I, are using the original lightcurves, whereas those corresponding to Model~II have had the slow varying component removed as described in Section~\ref{sec:CCF}. We fit to each year separately as the first year of observing was at a significantly lower cadence and contains large gaps in the $u'$, $B$, and $V$ lightcurves. This allows us to compare the ability to retrieve a delay spectrum for different intensities of AGN monitoring. It also allows comparison between \pyroa\ and CCF as the CCF method is unsuitable for the full 2 years as it linearly interpolates across the gap, which can distort the delay measurements.

\begin{table*}

  \caption{\pg\ Continuum Time Delay Measurements}
  \label{tab:Results}

    \def\arraystretch{1.3}
    \setlength{\tabcolsep}{8pt}
    \begin{threeparttable}
  \begin{tabular}{cccccccccc}
  
    \hline
  \multicolumn{4}{c}{}  & \multicolumn{3}{c}{Year 1}  & \multicolumn{3}{c}{Year 2} \\
\cmidrule(r){1-4}
\cmidrule(r){5-7}
\cmidrule(r){8-10}

\multicolumn{4}{c}{Photometric Bands} & \multicolumn{2}{c}{\pyroa}  & \multicolumn{1}{c}{CCF } & \multicolumn{1}{c}{\pyroa}  & \multicolumn{1}{c}{CCF } & \multicolumn{1}{c}{{\sc cream}}\\
\cmidrule(r){1-4}
\cmidrule(r){5-6}
\cmidrule(r){7-7}
\cmidrule(r){8-8}
\cmidrule(r){9-9}
\cmidrule(r){10-10}

    Filter & $\lambda_{\textrm{cent}}$ & $\lambda_{\textrm{width}}$ & Epochs & Model$^\dagger$ &  $\tau_i$ & $\tau_{\textrm{CCF}}$&  $\tau_i$ & $\tau_{\textrm{CCF}}$ & $\left< \tau \right>$ \\
    & (\r{A}) & (\r{A}) &  & & (Days)  & (Days) & (Days)  & (Days)& (Days) \\
    \hline
    
    $u'$ & 3540 & 570 & 398 & I & $-1.99^{+0.28}_{-0.29}$ &$-4.36^{+0.63}_{-0.72}$  & $-0.89^{+0.11}_{-0.12}$    & $-0.69^{+0.43}_{-0.46}$ & $-1.11 \pm 0.16$ \\
   & && & II &  &  & $-1.10^{+0.13}_{-0.13}$&  $-1.11^{+0.37}_{-0.37}$\\
    & && & III &  &  & $-1.21^{+0.11}_{-0.10}$&  \\
     \hline
  
    $B$ & 4361 & 890 & 864 & I   & $-1.10^{+0.18}_{-0.18}$&$-2.48^{+0.71}_{-0.71}$  &  $-0.53^{+0.07}_{-0.07}$  & $-0.84^{+0.19}_{-0.19}$ & $-0.34 \pm 0.19$\\
   & && & II &   & & $-0.43^{+0.06}_{-0.06}$ &  $-0.64^{+0.18}_{-0.16}$\\
    & && & III &  &  & $-0.42^{+0.06}_{-0.06}$&  \\

       \hline

    $g'$ & 4770 & 1500 & 1064 &I   &$0.00$ & $0.00^{+0.41}_{-0.41}$ & $0.00$  & $0.00^{+0.19}_{-0.20}$ & $0.00 \pm 0.20$ \\
   & && & II   &  &&  $0.00$ & $0.00^{+0.15}_{-0.15}$  \\
    & && & III &  &  & $0.00$&  \\

       \hline

    $V$ & 5468 & 840 & 856 &I & $0.31^{+0.18}_{-0.18}$ & $-0.27^{+0.77}_{-0.89}$ & $0.57^{+0.05}_{-0.05}$  & $0.93^{+0.16}_{-0.20}$ & $0.63 \pm 0.22$\\
   & && & II &    &&  $0.39^{+0.05}_{-0.05}$  &   $0.51^{+0.17}_{-0.17}$\\
    & && & III &  &  & $0.35^{+0.05}_{-0.05}$&  \\

       \hline

    $r'$ & 6215 & 1390 & 953 &I &$2.06^{+0.13}_{-0.14}$ &$1.53^{+0.48}_{-0.46}$  &  $1.77^{+0.06}_{-0.07}$ & $2.51^{+0.24}_{-0.27}$ & $1.33 \pm 0.24$\\
   & && & II &   & & $1.10^{+0.06}_{-0.06}$ &  $1.43^{+0.19}_{-0.20}$\\
    & && & III &  &  & $1.10^{+0.07}_{-0.07}$&  \\

       \hline

    $i'$ & 7545 & 1290 & 1008 & I & $2.82^{+0.18}_{-0.17}$& $2.58^{+0.72}_{-0.72}$ &   $2.06^{+0.10}_{-0.09}$  & $2.99^{+0.33}_{-0.27}$ & $2.64 \pm 0.29$ \\
   & && & II   &   &   &$1.39^{+0.09}_{-0.09}$ & $2.00^{+0.28}_{-0.29}$\\
& && & III &  &  & $1.51^{+0.10}_{-0.10}$&  \\

       \hline

    $z_s$ & 8700 & 1040 &	909 & I   & $3.27^{+0.24}_{-0.27}$ & $2.94^{+0.93}_{-0.99}$ &  $3.30^{+0.21}_{-0.19}$  & $4.74^{+0.41}_{-0.35}$ & $3.78 \pm 0.31$\\
   & && & II &  & & $2.61^{+0.19}_{-0.21}$ &  $3.63^{+0.51}_{-0.49}$\\
& && & III &  &  & $3.31^{+0.26}_{-0.27}$&  \\

    \hline
  \end{tabular}
\begin{tablenotes}
    \item[$\dagger$] Model~I does not account for a slow varying component, whereas Model~II does. In the case of CCF, this component is subtracted for Model~II, before calculation of the cross-correlation function. Model~III for \pyroa\ accounts for a slow varying component and different levels of blurring for each lightcurve. Lags for CREAM have been remeasured as well in reference to $g'$.
  \end{tablenotes}
  \end{threeparttable}
 \end{table*}

For the first year of data, \pyroa\ was only able to achieve a good fit for Model~I, where the lack of data prevented the more complex models from being constrained. In addition to the much sparser data, the $u'$, $B$, and $V$ bands had significantly less coverage than the other bands, preventing a reasonable delay from being obtained from the CCF method, while \pyroa\ provided more sensible results. The \pyroa\ results show the expected increasing delay with wavelength, however they are much more uncertain than the second year of data. The lack of data/clear shape prevents us from modelling any slow varying component. A plot of the \pyroa\ fit for the first year of data is shown in Fig.~\ref{fig:Year1}

\begin{figure*}
\hspace*{-0.8cm}                                                           
    \centering
	\includegraphics[width=19cm]{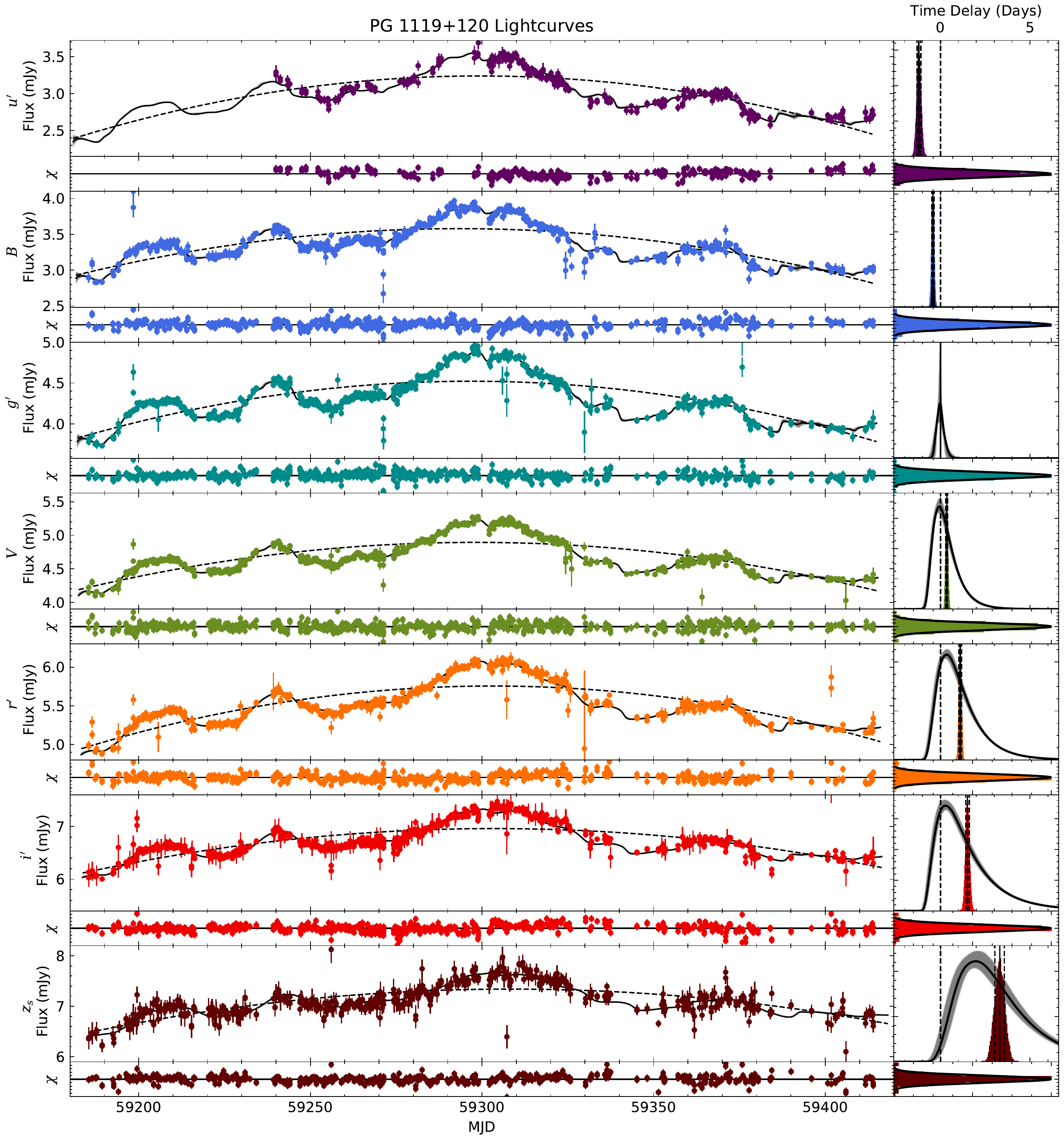}
    \caption{LCO lightcurves overlaid with the \pyroa\ Model~III (black line) and its error envelope in grey. The slow varying component given by Eqn.~(\ref{eqn:SlowComp}), is shown with a dashed line. The right hand panels show the posterior distributions for the time delay parameters with the median and 1$\sigma$ confidence intervals displayed on the histograms. Also on these panels are the log-Gaussian delay distributions, $\Psi_i(\tau)$, in black with their $1\sigma$ error envelopes. The normalised residuals, $\chi$, for each lightcurve are also shown between -5$\sigma$ and 5$\sigma$, with the colour corresponding to the appropriate lightcurve. The right panels of the residuals show a histogram of those normalised residuals, in comparison with the expected Gaussian distribution in black.}
    \label{fig:PGFit2}
\end{figure*}

For the second year of data, \pyroa\ was able to achieve a good fit for all models, constraining the delays with small uncertainties. Similarly, the CCF found delays that increased with wavelength for both the original lightcurves, and when the slow varying component had been removed. However, the CCF and the different models of the \pyroa\ fit show some disagreement which we investigate further. The smaller uncertainties on the delays and ability to investigate the slow varying component prompts us to focus on the second year results in our analysis of the accretion disc. Plots for model's I and II, are shown in Fig.\ref{fig:PGFitModelI} and \ref{fig:PGFitModelII} respectively. The Model~III fit is shown in Fig.~\ref{fig:PGFit2}.

\begin{table}
\centering
  \caption{\pyroa\ Lightcurve Model Comparison}
  \label{tab:Comparison}
    \def\arraystretch{1.2}
    \setlength{\tabcolsep}{6pt}
    \begin{threeparttable}
  \begin{tabular}{cccccccc}
  
    \hline

    Model & $\chi^2$  & $\chi_{\nu}^2$ & $P\ln N$ & $\sum \ln \sigma^2$ & BIC  \\
    \hline
    I & 4866.0 & 1.14 & 785.5 & 1825.0 & 7476.5 \\
    II & 4852.2 & 1.14 & 957.2 & 1158.4 & 6967.8 \\
    III & 4905.0 & 1.15 & 1019.1 & 1021.1 & 6945.2\\ 

    \hline
  
  \end{tabular}
\begin{tablenotes}
    \item[] $\chi_{\nu}^2$ = $\frac{\chi^2}{N-P}$ is the reduced $\chi^2$ for $N$ data points and model with $P$ parameters.

  \end{tablenotes}
  \end{threeparttable}
 \end{table}

To compare the goodness of fit of the two \pyroa\  models, we investigate the difference in the Bayesian Information Criterion (BIC) and its constituent components. The BIC is a ``badness of fit'' statistic, minimised by the simplest best fitting model. It is comprised of the $\chi^2$ statistic which measures how close the model is to the data, $P\ln N$ which is a penalty that scales with the number of parameters, $P$, for a given number of data points, $N$, and $\sum \ln \sigma^2$, which is a penalty that scales with the size of the error bars for the flux measurements. The exact relationship is given by equations (11, 12, 13) in \citet{Donnan2021}. The values of these statistics are shown in Table~\ref{tab:Comparison} for each model.

Between models I and II, the $\chi^2$ value reduces as well as $\sum \ln \sigma^2$, leading to a much lower BIC despite the increase number of parameters. The lower BIC is largely driven by the error bars being expanded less, which suggests the lightcurves are better modelled by a single driving lightcurve shifted in time, after the slow variations are removed. This suggests that the slow component acts on a different timescale and is investigated further in Section~\ref{sec:diffuseBLR}. Therefore including a slow varying component improves the fit to the lightcurves. Model~III shows a smaller improvement to the BIC over Model~II. Again the error bars are expanded less as the different levels of smoothing for each lightcurve provides a better fit, particular for the very short and very long wavelength lightcurves. As Model~III has the lowest BIC, we use its results primarily in the remaining analysis.

\begin{figure*}
	\includegraphics[width=18cm]{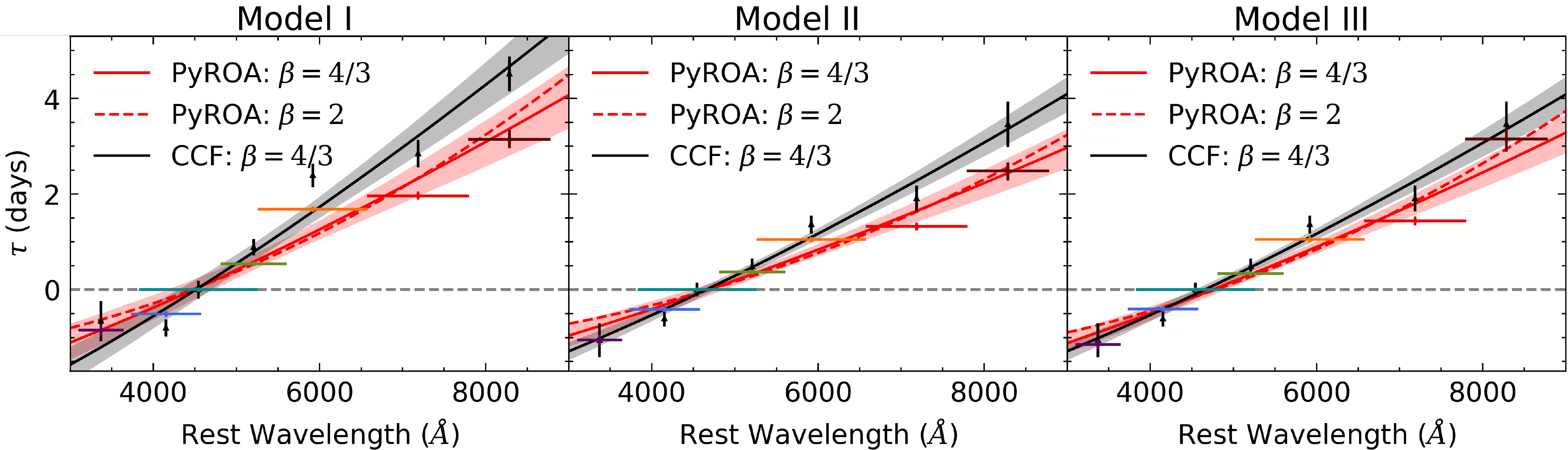}
    \caption{Delay spectra for each model as given in Table~\ref{tab:Results}, after shifting to the rest frame. The \pyroa\ delays are shown with the coloured points, whereas the CCF delays are shown with the black triangles. The width of the filters are shown as an error bar in the x-direction. The CCF delays shown for Model~I  are for the original lightcurves, whereas for models II and III, a slow varying component was removed. For each the model given by Eqn.~(\ref{eqn:delay}) is shown for a fixed $\beta = 4/3$, fitted to the \pyroa\ and CCF delays separately, where the best fit parameters are given in Table~\ref{tab:DelaySpecFits}. Additionally, the fit with a fixed $\beta = 2$ are shown for the \pyroa\ delays in the dashed red line. }
    \label{fig:DelaySpec} 
\end{figure*}

The resulting delay spectra are shown in Fig.~\ref{fig:DelaySpec}, after shifting to rest frame wavelength and removing cosmic time dilation by dividing the delays by $(1+z)$. To test the accretion structure, we fit a simple power law, given by
\begin{equation}
\label{eqn:delay}
    \tau = \tau_0 \left[ \left(\frac{\lambda}{\lambda_0} \right)^{\beta} - y_0\right],
\end{equation}
where $\tau_0$ is the amplitude, $\beta$ is the power, $\lambda_0 = \frac{4770}{1+z}$ \AA\,  is the reference wavelength, and $y_0$ allows the wavelength of zero time delay to vary from $\lambda_0$. We fit this model for a fixed $\beta = 4/3$ to both the \pyroa\ results and CCF results separately. Additionally, we fit the model for a fixed $\beta = 2$ to the \pyroa\ results. We also fit the {\sc cream} delays relative to the $g'$ band, to measure the amplitude, $\tau_0$, as the {\sc cream} delays already assumes the 4/3 delay spectrum.

We used {\sc emcee} \citep{Foreman-Mackey2013} to perform MCMC sampling, including a noise model to scale the errors of the delays by a fractional amount, $e^{2 \ln f}$. We used 20{,}000 samples, 32 walkers and discarded the initial 15{,}000 as burn-in. The best fit parameters are reported in Table~\ref{tab:DelaySpecFits}, where we calculate the reduced $\chi^2$ and the log-likelihood/d.o.f to evaluate the model's goodness of fit. We also fit this model to the CREAM delays measured relative to the $g'$ band in order to compare the amplitude with \pyroa\ and CCF.

We find that the resulting delay spectrum for both \pyroa\ and the CCF are broadly consistent with a power of $4/3$ and thus a temperature profile of $T \propto R^{-3/4}$. In both cases we also find that the $4/3$ power law provides a better fit after accounting for the slow variations, which again suggests that these arise on a different timescale that the disc reverberation. The power of $\beta=2$ is not inconsistent with the data however in each case provides a lower log-likelihood than the $4/3$ fit, and thus stronger evidence towards the temperature profile of $T \propto R^{-3/4}$. We find the largest log-likelihood for the \pyroa\ Model~III results which is encouraging as this lightcurve model also provided the best fit to the lightcurve data. 

For the case where the slow variations were not accounted for (Model~I), the $u'$ shows excess delay compared to model delay spectrum. We attribute this to diffuse continuum from the BLR as explained in Section~\ref{sec:diffuseBLR}. This effect also explains the larger amplitude for \pyroa\ I vs II and III, as the former is distorted with emission from larger radii which the BLR is though to be at compared to the accretion disc \citep[e.g ][]{Netzer2015}.

We also see differences in the amplitude of the delay spectra between \pyroa\ and the CCF where \pyroa\ finds and amplitude $\sim 2\sigma$ lower. The theoretical transfer function for accretion disc reprocessing is asymmetric around the mean delay \citep[e.g. ][]{Cackett2007, Starkey2016}, and therefore treating the delays symmetrically can introduce a bias. Specifically \citet{Chan2020} showed that modelling the lightcurves using this transfer function, such as CREAM does, finds delays $\sim 30 \%$ larger than the CCF and JAVELIN \citep[][]{JAVELIN}, the latter two treating the delays symmetrically. This can explain the larger amplitude of the delay spectrum for CREAM compared to the CCF and \pyroa\, where $\tau_0$ is $\sim 12 \%$ larger than the CCF II results (as the CREAM fit was to the detrended lightcurves) and $\sim 53 \%$ larger than the \pyroa\ II results. The lower amplitude for \pyroa\ compared to CCF may also be due to this effect. \pyroa\ models the lightcurves with a single level of smoothing, which is more sensitive to shorter delays at long wavelengths and longer delays at short wavelengths. Once different levels of smoothing were introduced into \pyroa\ for Model~III, we see the expected increase in the amplitude of the delay spectrum compared to Model II, but only $\sim 13 \% $ larger. Future work is required to further investigate this effect, which is particularly important when drawing physical conclusions about the size of the accretion disc, although fortunately does not effect the relationship with wavelength used to probe the temperature profile.

\begin{table}
\centering
  \caption{Delay Spectrum Model Parameters}
  \label{tab:DelaySpecFits}
    \def\arraystretch{1.4}
    \setlength{\tabcolsep}{3pt}
    \begin{threeparttable}
  \begin{tabular}{ccccccccccc}
  
    \hline
       Lightcurve & $\beta$ & $\tau_0$ & $y_0$  & $\nu$ & $\ln f ^\ddagger$ & $\chi^2_{\nu}$ & $\ln L / \nu^\dagger$ \\
         Model   & &(Days) & & & &\\
    \hline
    \pyroa\ I & $4/3 ^{*}$ &$2.68^{+0.47}_{-0.56}$ & $0.98^{+0.07}_{-0.10}$  & 3&$0.86^{+0.23}_{-0.17}$& 1.88 & 0.98\\
     & $2^{*}$ &$1.52^{+0.32}_{-0.38}$ & $0.96^{+0.14}_{-0.21}$  & 3 & $0.94^{+0.21}_{-0.17}$& 1.08 & 0.66\\

    \pyroa\ II & $4/3 ^{*}$ &$2.05^{+0.30}_{-0.34}$ & $1.04^{+0.06}_{-0.07}$  & 3 &$0.65^{+0.24}_{-0.17}$& 1.12 & 1.84\\
     & $2^{*}$ &$1.13^{+0.22}_{-0.25}$ & $1.06^{+0.13}_{-0.16}$  & 3 & $0.76^{+0.23}_{-0.17}$& 1.11 & 1.39\\

    \pyroa\ III & $4/3 ^{*}$ &$2.32^{+0.29}_{-0.32}$ & $1.06^{+0.05}_{-0.05}$  & 3 & $0.59^{+0.22}_{-0.17}$& 1.13 & 1.94\\
     & $2^{*}$ &$1.32^{+0.21}_{-0.24}$ & $1.11^{+0.10}_{-0.11}$  & 3 & $0.71^{+0.21}_{-0.17}$& 1.08 & 1.48\\

    CCF I & $4/3 ^{*}$ &$3.77^{+0.48}_{-0.50}$ & $0.99^{+0.05}_{-0.06}$  & 4 & $0.37^{+0.18}_{-0.15}$& 1.09 & 0.69\\
    CCF II & $4/3 ^{*}$ &$2.82^{+0.26}_{-0.29}$ &$1.03^{+0.03}_{-0.04}$  & 4 & $0.08^{+0.19}_{-0.15}$& 1.08 & 1.77\\
    CREAM & $4/3 ^{*}$ &$3.15^{+0.04}_{-0.04}$ & $1.01^{+0.01}_{-0.01}$ & 4 & $-0.74^{+0.19}_{-0.15}$& 1.10 & 4.64 \\

    \hline
  
  \end{tabular}
\begin{tablenotes}
    \item[$\dagger$] $\ln L / \nu$ is the log-likelihood/d.o.f
    \item[$*$] Values of $\beta$ are fixed 
    \item[$\ddagger$] Log fractional error of original delay errors where errors are expanded by a factor of $e^{2 \ln f}$

  \end{tablenotes}
  \end{threeparttable}
 \end{table}


\subsection{Accretion disc modelling with CREAM}
\label{sec:cream}
The high cadence and S/N of the light curves are suitable to perform detailed analysis and infer physical properties of the accretion flows surrounding the supermassive black holes. We employed the AGN reverberation code {\sc cream} \citep{Starkey2016} to model the light curves within the standard lamp-post model scenario \citep{Haardt1991}. Here, we briefly describe the method, and we refer the reader to \citet{Starkey2016,Starkey2017} for further details on the implementation.

The model assumes that a single driving light curve $X(t)$ is responsible for the variability by displacing it in time ($\tau_{\rm CREAM}$), and applying a flux scaling (multiplicative factor) and shift (additive term) to match the light curve at each wavelength. The driving light curve is modelled as a Fourier series with a prior set to a random walk process (REF) and the model is fitted to all light curves simultaneously by convolving the driving light curve with the transfer function, $\Psi(\tau | \lambda)$. This transfer function is designed to capture the geometry effects of the accretion flow, which is a function of inclination angle, the product $\dot{M}\mbh$, and the radial temperature profile. Hence, this method permits inferences on physical quantities within the context of the disc reprocessing model, since the delay scales as $\langle\tau\rangle \propto (\dot{M}\mbh)^{1/3}\lambda^{4/3}$. Thus, we can fit for $\dot{M}\mbh$ given by the average delay of the individual light curves and calculate the average delay as the mean of the transfer function (which by construction the lag spectrum follows the $\tau_{\rm CREAM}\propto\lambda^{4/3}$ relationship). 
Given that {\sc cream} assumes that variability is solely driven by disc reprocessing, we used the data after removing the slowly variable component as measured by \pyroa\ Model~II (see Sec.~\ref{sec:DelaySpec} for details).

\begin{figure}
\centering
	\includegraphics[trim=0cm .1cm 0.2cm 0cm,clip,width=0.9\columnwidth]{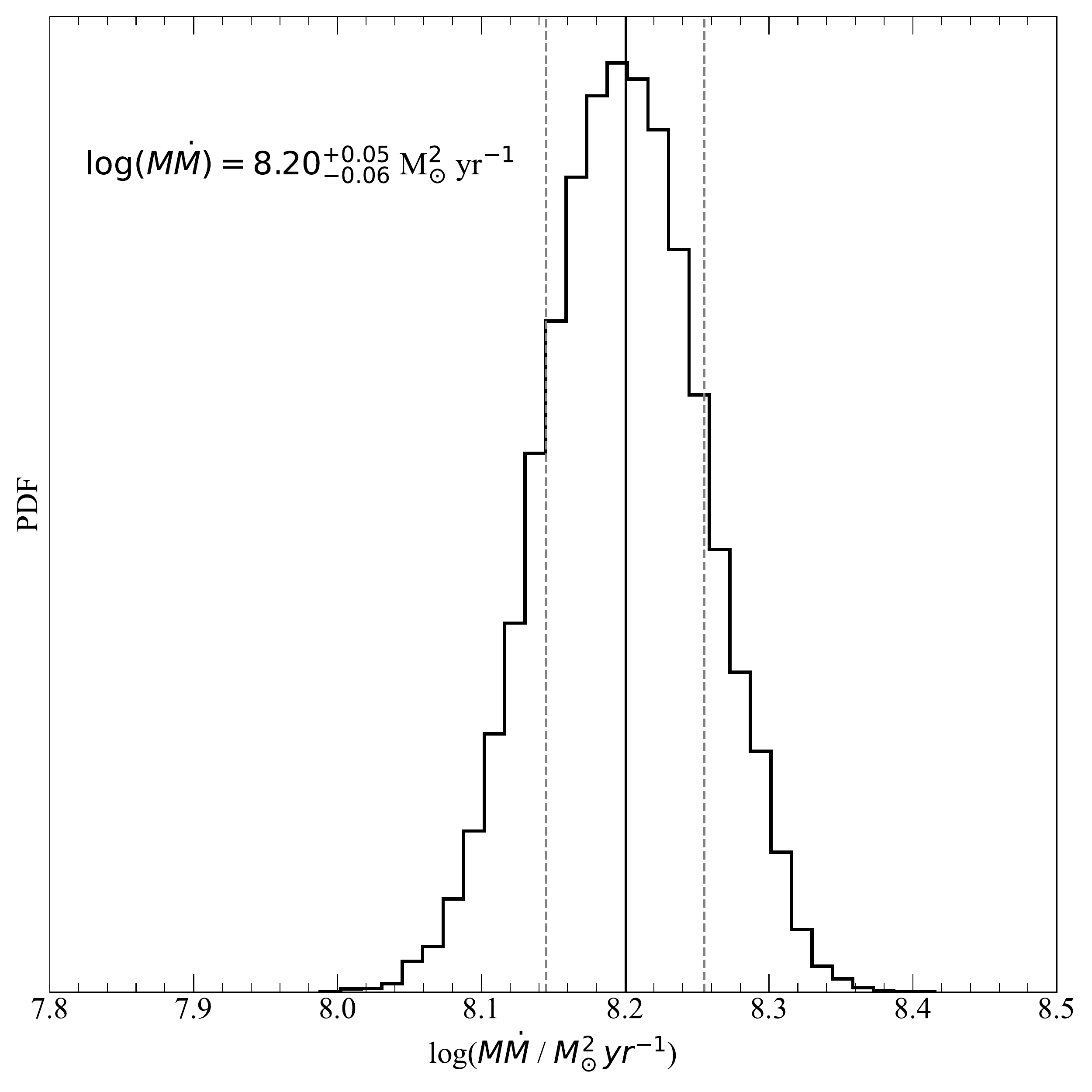}
    \caption{The marginalised posterior distribution for the $M\dot{M}$ as measured by {\sc cream}. The median value and its 68\% confidence interval are shown as the solid and dashed vertical lines, respectively.}
    \label{fig:cream_mdot} 
\end{figure}

\begin{figure*}
\hspace*{-0.8cm}                                                           
    \centering
	\includegraphics[trim=1cm 6.7cm 2cm 0cm,clip,width=18cm]{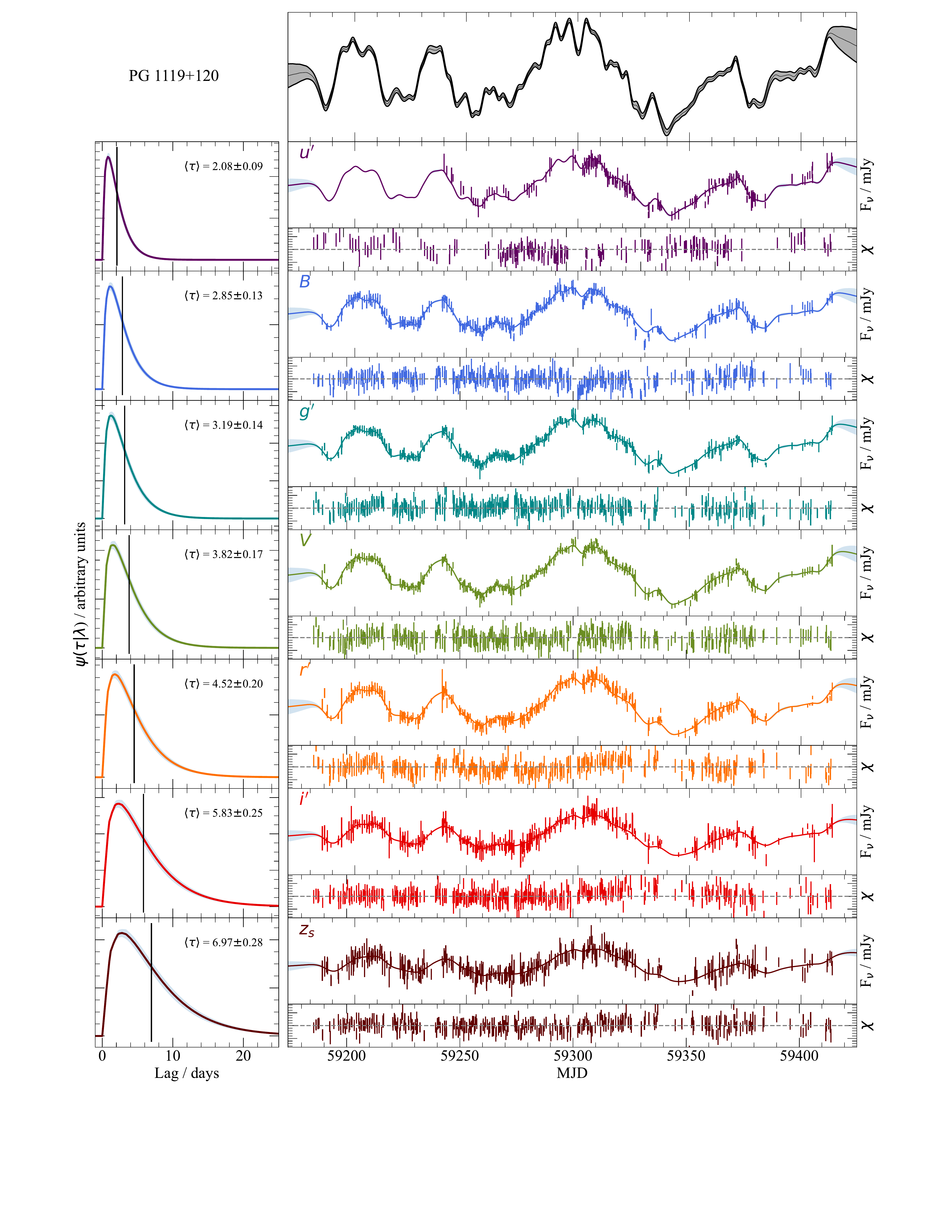}
    \caption{Reverberation model fit with {\sc cream} to a face-on accretion disc and a $T\propto R^{-3/4}$ temperature profile for \pg. Top panel shows the driving light curve. Left panels show the delay distribution for each band and the vertical black line shows the mean delay $\langle\tau\rangle$. Right panel shows the photometry for each band and the best fit of the driving light curve scaled to each band. Below each lightcurve, the sub-panel show the normalised residuals with the dotted line at $\chi=0$ for reference. All grey envelopes represent the 1$\sigma$ confidence interval. }
    \label{fig:creamfit}
\end{figure*}

Our model fit assumes a face-on disc ($i = 0$), since the effect of inclination angle on the measurement of the delay spectrum has been shown to be very small \citep{Starkey2016}. The model also assumes a standard disc temperature profile, $T\propto R^{-3/4}$. The noise model contains an additional scatter term added in quadrature for each lightcurve. The driving lightcurve has a fixed maximum Fourier frequency of $\nu_\mathrm{max} = 0.5$ cycles/day, set by the average time separation between observations in the LCO bands. 
The {\sc cream} fit for each AGN was done by exploring the posterior distribution of the model parameters with a Monte Carlo procedure, performing 150{,}000 iterations and discarding 50{,}000 as burn-in. We find a best value for  $\log(\dot{M}\mbh /\, {\rm M}_{\odot}^2\, {\rm yr}^{-1})=8.20^{+0.05}_{-0.06}$, shown in Fig.~\ref{fig:cream_mdot}. The fits to the individual lightcurves, their residuals as well as the inferred transfer functions, are shown in Fig.~\ref{fig:creamfit}.

\subsection{Spectral Energy Distribution}\label{sec:sed}
\begin{figure*}
	\includegraphics[width=18cm]{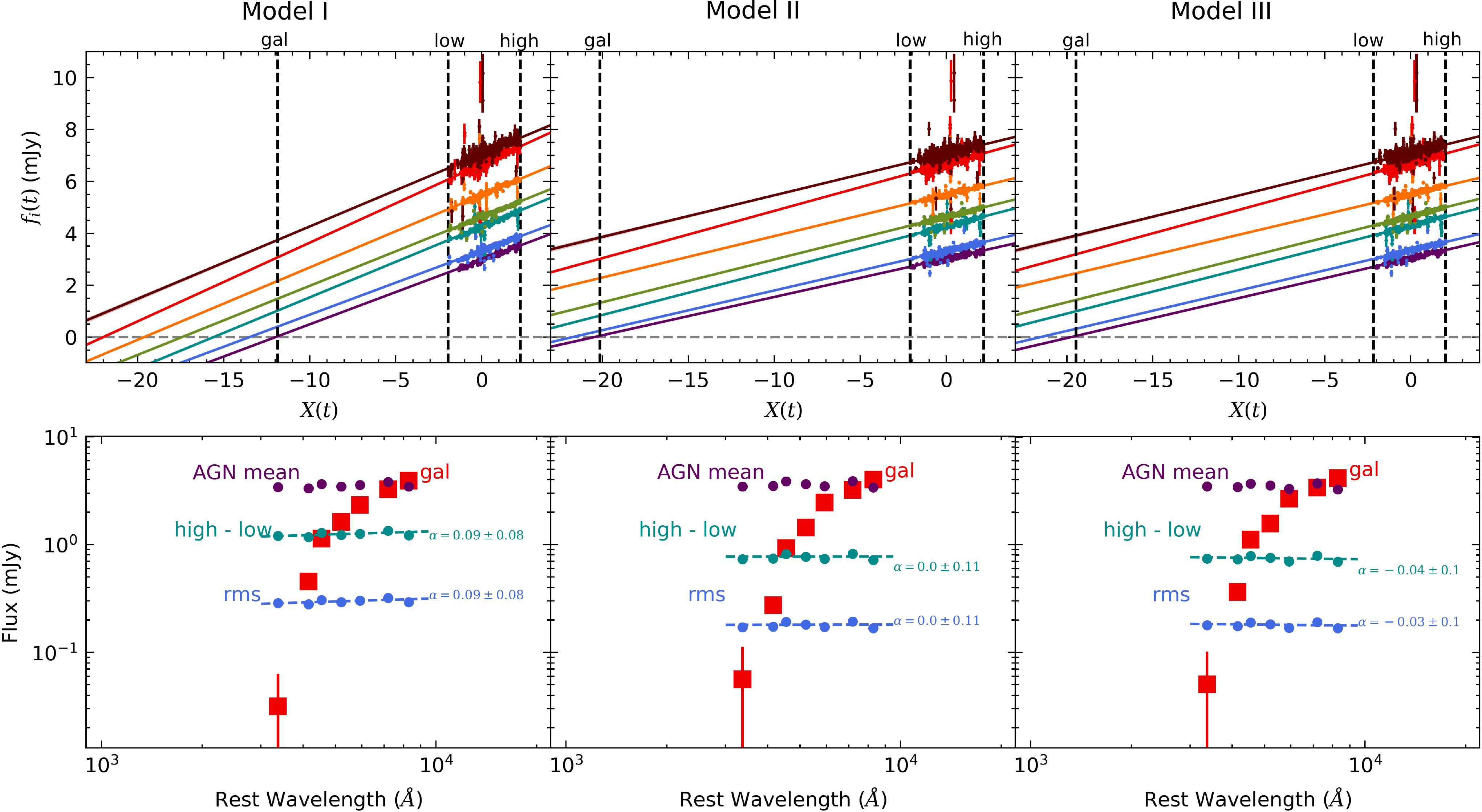}
    \caption{{\it Top:} Flux-flux plots for each \pyroa\ model where the flux for each band is plotted against the driving lightcurve, with the model fit as a straight line. The labels, gal, low, high, mark the values of the driving lightcurve for the host galaxy, the minimum flux and the maximum flux respectively. {\it Bottom:} Spectral energy distributions (SED), constructed from the various components of the photometric lightcurves. The AGN high-low, and rms fluxes are fitted with a power law, with power, $\alpha$, which is shown. The AGN mean is calculated from the mean values with the host galaxy contribution removed. The SED has been de-reddened using a MW dust law with E(B-V)=0.033.}
    \label{fig:SEDFig} 
\end{figure*}

From the modelling of the photometric lightcurves, we have obtained models for the driving lightcurve from \pyroa, $X(t)$, which describes the shape of the variability. This allows a decomposition of the photometric flux into AGN and host galaxy, which can be used to construct a spectral energy distribution. To do this, we construct a flux-flux plot where the flux of each band is plotted against the value of $X(t)$. For models including a slow component, this is subtracted from the flux before constructing. From Eqn.~(\ref{eqn:model1}), this forms a linear relationship where the gradient is provided by the rms parameters, $A_i$ and the intercept by the mean $B_i$. By extrapolating this straight line until the shortest wavelength band reaches zero flux within 1$\sigma$, we ``turn off'' the variability to remove the variable component of the flux. Taking the value of $X(t)$ at this point and reading of the flux values, provides an estimate of the host galaxy contribution. The high and low fluxes shown in Fig.~\ref{fig:SEDFig}, are the values where the driving lightcurve was at its maximum and minimum over the observed lightcurve.

Before constructing a SED, we de-redden the flux values to account for dust in the Milky Way using the model from \citealt{Fitzpatrick1999} and a dust extinction of E(B-V) = 0.033 \citep{Schlafly2011} before being shifted to the rest-frame.

The AGN contribution can be estimated in a few ways. Firstly we plot the rms flux and the range of flux (high - low), which shows how the amplitude of the variability changes with wavelength, and is independent of the host galaxy component. In order to compare with the theoretical power law of a thin accretion disc we fit the following model
\begin{equation}
\label{eqn:flaw}
    f_{\nu} = f_0 \lambda^{\alpha},
\end{equation}
where $f_0$ is the amplitude and $\alpha$ is the power. The best fit values of $\alpha$, are shown on Fig.~\ref{fig:SEDFig}. To obtain an estimation of the AGN flux we subtract the host galaxy component from the mean fluxes which are shown in Fig.~\ref{fig:SEDFig} as AGN mean.

\begin{figure}
	\includegraphics[width=\columnwidth]{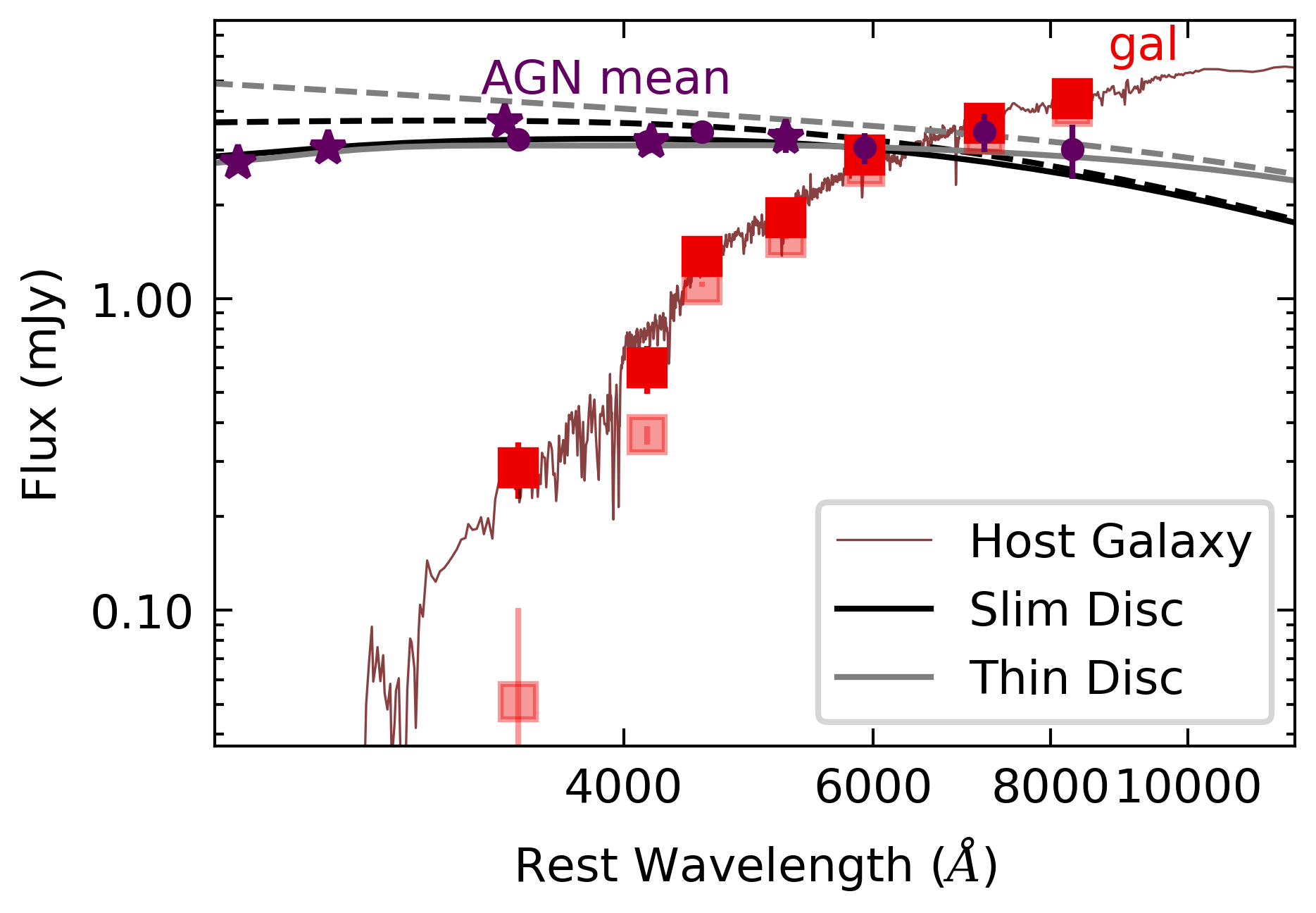}
    \caption{SED showing the host galaxy and AGN contribution from Model~III after correcting the host galaxy contribution using the best fitting template spectrum from \citet{Brown2014} (NGC 4550) as shown. The host galaxy contribution after fitting to the template spectrum are shown in the solid red squares, whereas the original fluxes are shown as the transparent red squares. The mean AGN flux is shown in the purple circles for LCO and purple stars for Swift measurements. Overlayed are the fitted disc models for a thin disk (grey) and a slim disc (black) from  \citet{Kubota2018, Kubota2019}, where the parameters are shown in Table~\ref{tab:DiscModel}. The solid lines show the models with dust extinction while the dashed shows without extinction.}
    \label{fig:SEDNew} 
\end{figure}

Typically this kind of spectral decomposition uses UV bands to determine the value of $X(t)$ for the host galaxy contribution \citep[e.g][]{Cackett2020, HernandezSantisteban2020}, where the shortest wavelength UV band reaches zero flux, however our analysis uses only optical bands. Using the condition that $u'$ band reaches zero flux likely underestimates the host galaxy component with a value of $X(t)$ that is too small, as the the host galaxy still likely emits significantly at this wavelength. To improve this we fit 128 calibrated template spectra from \citet{Brown2014} which covers a range of galaxy morphologies, colour and star formation rates. We scale each template spectra using the Model~III flux-flux plot and allow the value of $X(t)$ for the host galaxy contribution to vary as a free parameter using {\sc emcee} \citep{Foreman-Mackey2013},  including a noise model to scale the galaxy flux errors. For each band, the template spectra are convolved with a top-hat function of the width of each filter before scaling to fit the data. We compute the relative log-likelihood of each template fit to assess which template provides the best fit to our data. We find the best fit template to be that of NGC 4550 with $X_{\textrm{gal}}(t) = -18.16^{+0.34}_{-0.39}$, which is an elliptical galaxy hosting an old stellar population. All the best fitting templates were similar - very red due to an old stellar population. \pg\ is known to be a disk galaxy \citep[][]{Shangguan2020} with very faint spiral arms containing star forming regions \citep[][]{Surace2001} and so the detection of an old stellar population suggests \pg \ may be close to a lenticular galaxy with few young stars and/or that the 5'' aperture used to extract the photometric lightcurves largely excludes the spiral arms. Taking $X_{\textrm{gal}}(t) = -18.16$, the resulting mean AGN flux and host galaxy flux are shown in Table~\ref{tab:SED} and plotted in Fig.~\ref{fig:SEDNew} with the template galaxy spectrum. 

\begin{table}
\centering
  \caption{Spectral Decomposition}
  \label{tab:SED}
    \def\arraystretch{1.2}
    \setlength{\tabcolsep}{8pt}
    \begin{threeparttable}
  \begin{tabular}{cccc}
  
    \hline
    $\lambda_{\textrm{rest}}$ & Host Galaxy & AGN Mean & Telescope  \\
    (\AA) & (mJy) & (mJy) \\
    \hline
    1835.8 & $0.0114 \pm 0.0017$ & $2.80 \pm 0.18$ & Swift\\
    2139.4 & $0.0165 \pm 0.0025$ & $2.73 \pm 0.18$ & Swift\\
    2475.7 & $0.0512 \pm 0.0077$ & $3.05\pm 0.19$ & Swift\\
    3299.4 & $0.243 \pm 0.036$ & $3.70\pm 0.19$ & Swift\\
    4182.1 & $0.89 \pm 0.13$ & $3.19\pm 0.23$ & Swift\\
    5206.6 & $1.94 \pm 0.29$ & $3.29\pm 0.36$ & Swift\\
    
    3370.8 & $0.287 \pm 0.059$ & $3.239 \pm 0.059$ & LCO\\
    4152.5 & $0.60 \pm 0.10$ & $3.19 \pm 0.10$& LCO\\
    4542.0 & $1.37 \pm 0.16$ & $3.43 \pm 0.16$& LCO\\    
    5206.6 & $1.82 \pm 0.26$ & $3.29 \pm 0.26$& LCO\\    
    5917.9 & $2.90 \pm 0.35$ & $3.06 \pm 0.35$& LCO\\    
    7194.3 & $3.65 \pm 0.49$ & $3.45 \pm 0.49$& LCO\\ 
    8284.1 & $4.38 \pm 0.58$ & $3.04 \pm 0.59$& LCO\\

    \hline
  
  \end{tabular}
\begin{tablenotes}
   \item[] A SED of these values is plotted in Fig.~\ref{fig:SEDNew}.
  \end{tablenotes}
  \end{threeparttable}
 \end{table}

\begin{figure}
	\includegraphics[width=\columnwidth]{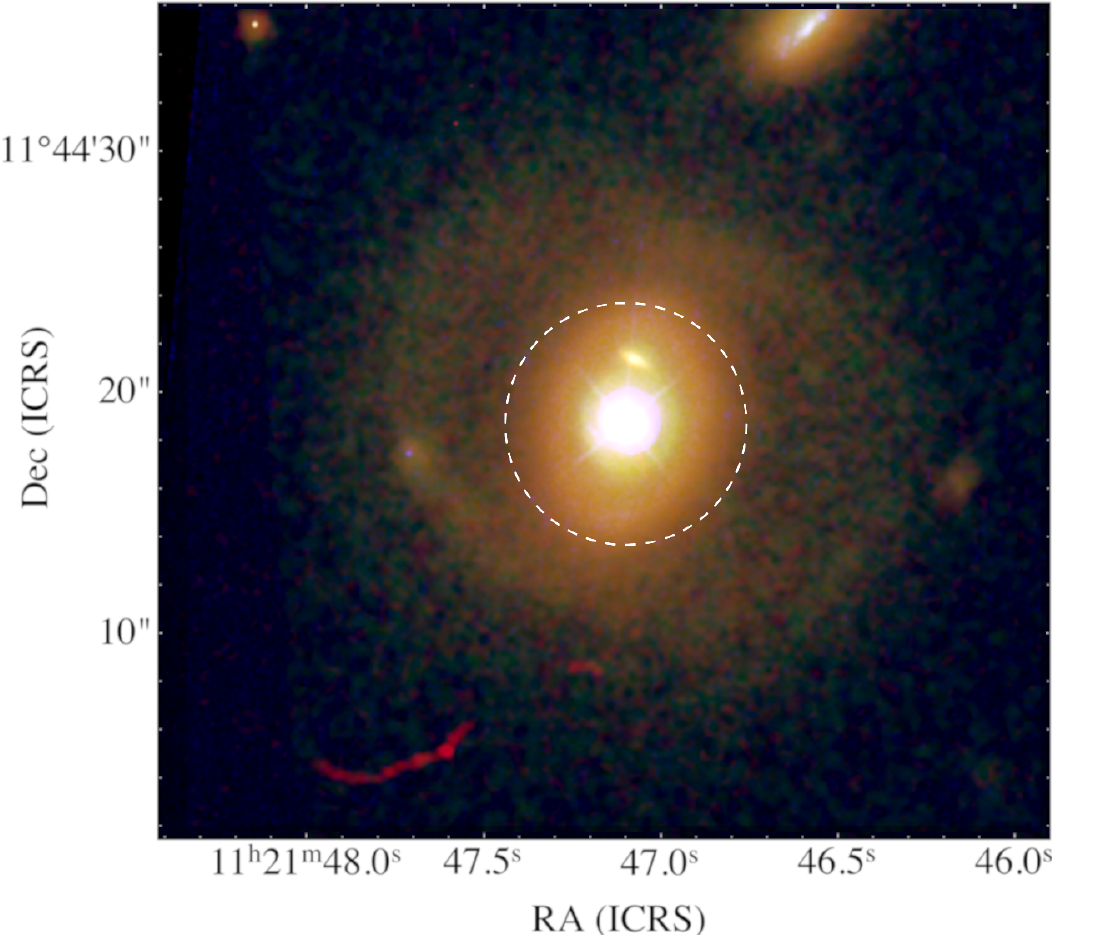}
    \caption{Colour composite image of \pg using the F438W, F814W, F160W HST filters from \citep[][]{Shangguan2020}  and $g$, $r$ and $z$ DECaLS filters \citep[][]{DECALS2019}. As the outer spiral arms are faint in the HST image we use the DECaLS imaging to provide the diffuse emission although at a lower resolution. The 5'' aperture used to extract the lightcurves is shown with a dashed white circle. There is also a red compact source north of the central galactic bulge.}
    \label{fig:Image} 
\end{figure}

As a sanity check we compared the 5'' aperture to an image of the galaxy. In Fig.~\ref{fig:Image} we show a colour composite image of PG 1119+120 using the F438W, F814W, F160W filters from Hubble Space Telescope (HST) data \citep[][]{Shangguan2020} and $g$, $r$ and $z$ band images from The Dark Energy Camera Legacy Survey (DECaLS)
\citep[][]{DECALS2019}. From this image, the aperture excludes much of the star forming regions in the faint spiral arms of \pg. This confirms the detection of an old and red stellar population from our best fitting template spectrum. We also note the bright region to the north of the nucleus. This region is contributing flux to the lightcurves and it is unclear whether it is a star forming region within the galaxy or a background galaxy. This region is very red \citep[][]{Surace2001}, which suggests high dust extinction in the case of a starburst or a background galaxy redshifted and reddened by the ISM dust.

While the Swift observations are too sparse ($\sim$ 3 epochs) to measure any delays or decompose through flux-flux analysis, we can estimate the AGN component using this host galaxy template and the sample mean of the raw data. We first calculate the optimal average of each of the Swift lightcurves and estimate the error. We then convolve the host galaxy template with the bandpass for each filter and subtract from the mean to give an estimate of the AGN component. As this mean is determined from three epochs, it is likely considerably offset from the true mean as the AGN may have been dimmer/brighter than average at these specific epochs. Therefore we correct the calculated mean by calculating the offset between the Swift B and V bands and the interpolated LCO measurements. We estimate the error in these fluxes by propagating the error in the sample mean, the noise model from the galaxy template and the offset. The resulting fluxes are given in Table~\ref{tab:SED} and are plotted in Fig.~\ref{fig:SEDNew}.


\section{Discussion}
\label{sec:Discuss}

\subsection{Spectral Shape}
\label{sec:SpecShape}
From Section~\ref{sec:sed} we found the flux of the AGN component to be significantly flatter than expected for a thin accretion disc. We found the delay spectrum to be consistent with $\tau(\lambda) \propto \lambda^{4/3}$ for a thin disc \citep{Shakura1973} which suggests a temperature profile of $T(R) \propto R^{-3/4}$. From this temperature profile the expected flux should follow $f_{\nu} \propto \lambda^{-1/3}$ however we find a flatter spectrum with $f_{\nu} \sim \textrm{constant}$.

Our spectral slope is too red compared to the thin disc power law but not red enough to be consistent with the slim disc $f_{\nu} \propto \lambda$ within the photon trapping radius, therefore we suspect a thin disc spectrum reddened by  dust in the ISM of the host galaxy as we have only accounted for foreground MW dust extinction. Dust reddening is a common feature in AGN SED's \citep[e.g. ][]{Baron2016, Brown2019} and so is not intrinsically unexpected here, however most targets of RM studies show spectral slopes that are strongly consistent with $f_{\nu} \propto \lambda^{-1/3}$ \citep[e.g. ][]{Fausnaugh2016, Cackett2020, HernandezSantisteban2020}. This is likely due to selection bias effects where targets chosen for RM studies are highly variable and have inclinations close to face-on and therefore have a lack of dust along the line of sight towards the nucleus. Using a SMC extinction curve \citep[][]{Gordon2003} requires and E(B-V) $\gtrapprox 0.05$ to be consistent with $f_{\nu} \propto \lambda^{-1/3}$ within $1\sigma$. To get exactly $\alpha = 0.33$ requires an extinction of E(B-V) $\approx 0.07$. Both of these values are very plausible/realistic values for interstellar dust extinction towards AGN \citep[e.g ][]{Baron2016}. For the slim disc scenario, dust extinction would give a spectrum even redder than $f_{\nu} \propto \lambda$.


\subsection{Accretion Rate}
\label{sec:accretionrate}
Both the delay spectrum and flux spectrum of the AGN depend on the temperature profile of the disc, which scales with the accretion rate and black hole mass, therefore can be used to estimate the accretion rate. The theoretical temperature structure of a geometrically thin, optically thick accretion disc can be derived assuming viscous heating in the disc and passive heating from a variable irradiating source at height, $H_x$ above the central SMBH. This is given by 
\begin{equation}
    T (R) = \left[ \left(   \frac{3 G \mbh \dot{M}}{8 \pi \sigma R^3  } \right) + \left( \frac{(1-A) L_x H_x}{4 \pi \sigma R^3}\right)   \right]^{1/4}, 
\end{equation}
where 
$\dot{M}$ is the accretion rate, $R$ is the radius in the disc, $A$ is the disc albedo, $L_x$ is the luminosity of the variable source, $\sigma$ is the Stefan-Boltzmann constant, and $G$ is the gravitational constant \citep{Cackett2007}. Following \citealt{Fausnaugh2016}, the equation can be simplified by setting $ (1-A)\, L_x \,H_x/R = \kappa\, G \,\mbh\, \dot{M}/2 \,R$, where $\kappa$ is the local ratio of passive to internal heating. Therefore the temperature profile becomes:
\begin{equation} 
    \label{eqn:TempProfile}
    T(R)=T_0 \,R^{-3/4}, \quad T_0 = \left[ \left( \frac{G \,\mbh \,L_{\textrm{Edd}}}{8 \,\pi \,\sigma \,c^2} \right) (3 + \kappa) \,\dot{\mathscr{M}} \right]^{1/4}
\end{equation}
where $\dot{\mathscr{M}}$ is the dimensionless accretion rate, 
\begin{equation}
\label{eqn:AccRate}
    \dot{\mathscr{M}}\equiv\frac{\dot{M}\,c^2}{L_{\rm Edd}} = \frac{\lambda_{\rm Edd}}{\eta},
\end{equation}
where $\dot{M}$ is the accretion rate, $L_{\rm Edd}$ is the Eddington luminosity for a given black hole mass, which relates to the Eddington ratio $\lambda_{\rm Edd} = L_{\rm bol}/L_{\rm Edd}$ through the radiative efficiency, $\eta$. We use the dimensionless accretion rate as the radiative efficiency is expected to decrease as a function of the accretion rate \citep{Wang1999} in the case of a slim disc.

An initial estimate of the accretion rate can be calculated from the observed flux at 5100 \AA. Interpolating the photometric fluxes of the AGN component (Table~\ref{tab:SED}) yields a flux of $f_{5100} = 3.32$ mJy, which gives a luminosity of $L_{5100} = 1.24 \times 10^{44}$ erg s\textsuperscript{-1} using a luminosity distance of 230.6~Mpc. Following \citet{Du2015, Du2016}, the dimensionless accretion rate can be estimated from the luminosity through
\begin{equation}
\label{eqn:AccRate2}
    \dot{\mathscr{M}}= 20.1 \left( \frac{L_{5100}}{10^{44} \cos{i}}\right)^{3/2} M_{7}^{-2},
\end{equation}
where $M_7$ is the black hole mass in $10^7 M_{\odot}$, and $i$ is the inclination of the disc. Assuming the torus is aligned with the disc, we use the mean inclination for type-1 AGN of $\cos{i} = 0.75$ (the uncertainty in the mass will dominate the error budget here). This gives a dimensionless accretion rate of $\dot{\mathscr{M}} = 42.9$ for $\log \mbh/\msun = 7.0$. This is in excess of the threshold of $\dot{\mathscr{M}} = 3$ where objects with $\dot{\mathscr{M}} > 3$ show smaller H$\beta$ lags than lower accretion SMBHs \citep[e.g.][]{Du2016}. This suggests that \pg \ is accreting beyond the Eddington limit, where a slim disc would be expected within the photon trapping radius. The photon trapping radius depends linearly on the accretion rate \citep[][]{Du2016}, and can be estimated using
\begin{equation}
    R_{\rm trap} = 144 \left( \frac{\dot{\mathscr{M}}}{10^2} \right) R_{\rm Sch},
\end{equation}
where $R_{\rm Sch}$ is the Schwarzschild radius which depends only on the mass of the black hole. The accretion rate estimated gives a photon trapping radius of $R_{\rm trap} = 61.7 R_{\rm Sch}$. Using the temperature profile in Eqn.~(\ref{eqn:TempProfile}), at these radii a blackbody would peak at wavelengths $\sim 370$\AA \, which is in the EUV and significantly less than the shortest wavelength filter used here centred at $1835.8$\AA. Our UV/optical data likely probes radii of $\sim 4000 - 30000 R_{\rm Sch}$, therefore the presence of a slim disc may not contribute significantly to our observed delay/flux spectrum.


A better estimate of the accretion rate can be made modelling the entire SED with an accretion disc model. We model the SED using a model for a standard thin disc as well as a slim disc. For th thin disc we use the model of \citet{Kubota2018} which consists of an outer thin Novikov-Thorne disc with an inner warm Comptonizing region and a hot corona, implemented in python using the {\sc QSOSED} code\footnote{\url{https://github.com/arnauqb/qsosed}}. With this model the Eddington ratio is a free parameter, where the radiative efficiency is calculated from the innermost stable circular orbit. 

To model a slim disc, we modify the {\sc QSOSED} code to use the \citet{Kubota2019} slim disc model, which is identical in the sub-Eddington regime but varies at higher accretion rates. Specifically, the inner radius and thus the radiative efficiency now depends on $\lambda_{\rm Edd}$ according to Eqn.~(1) of \citet{Kubota2019} and the temperature structure changes according to their Eqn.~(2). We fit each model to the AGN mean fluxes for a fixed black hole mass of $\log \mbh/\msun = 7.0$ and allow the Eddington ratio ($\lambda_{\rm Edd}$), hard X-ray fraction ($X_{\rm{frac}}$) and black hole spin ($a$) to vary as free parameters. We also allow E(B-V) to vary as a free parameter, using a SMC extinction curve \citep[][]{Gordon2003}, to de-redden the AGN flux and allow a measurement of the possible dust extinction for \pg.

The results of fitting each model are shown in Table~\ref{tab:DiscModel}, where we show the median and 1$\sigma$ confidence intervals while Fig.~\ref{fig:SEDNew} shows the best fitting models. We also show the posterior probability distributions in Fig. \ref{fig:DiscPost}.

\begin{table}
\centering
  \caption{Accretion Disc SED Models}
  \label{tab:DiscModel}
    \def\arraystretch{1.2}
    \setlength{\tabcolsep}{8pt}
    \begin{threeparttable}
  \begin{tabular}{cccc}
  
    \hline
    Parameter & Thin Disc & Slim Disc  \\
    \hline
    $\lambda_{\rm{Edd}}$ & $15.9^{+7.4}_{-5.7}$ & $3.26^{+0.20}_{-0.17}$\\
    $\rm{E(B-V)}$ & $0.07^{+0.01}_{-0.01}$ & $0.03^{+0.01}_{-0.01}$\\
    $X_{\rm{frac}}$ &  $4.79^{+5.39}_{-3.38}$ & $1.13^{+1.10}_{-0.77}$\\
    $a$ & $0.86^{+0.11}_{-0.31}$ & $0.69^{+0.24}_{-0.36}$\\
    \hline
    $h$ & $7.8^{+8.5}_{-3.6} R_{g}$ & $10.5^{+17.5}_{-5.4} R_{g}$\\
    $\eta$ & $0.141^{+0.074}_{-0.054}$ & $0.103^{+0.075}_{-0.032}$\\
    $\chi^2_{\nu}{}^{*}$ & 5.3 & 1.27 \\

    \hline
  
  \end{tabular}
\begin{tablenotes}
   \item[] A plot of these models is shown in Fig.~\ref{fig:SEDNew}.
   \item[*] The reduced chi-squared is calculated for a model using the median parameters given above. As the posterior probability distributions are not Gaussian/symmetric, these may not be useful for comparing models.

  \end{tablenotes}
  \end{threeparttable}
 \end{table}

Both disc models provide a good fit to the photometry, however the slim disc model gives a slightly lower reduced chi-squared. The accretion rate is super-Eddington in both cases with the thin disc requiring a much higher accretion rate than the slim disc. The thin disc requires more dust to flatten the model to fit the data. In both cases the black hole spin is close to 1. Using the MCMC samples we also calculate posteriors for the corona height, $h$, and the radiative efficiency, $\eta$, for each model, and are also shown in Table~\ref{tab:DiscModel}. 

The delay spectrum can also be used to test the accretion rate as well as the size of the disc. From the temperature structure given by Eqn.~(\ref{eqn:TempProfile}), the delay spectrum is given by:
\begin{equation}
\label{eqn:deltatau}
    \Delta \tau = \tau_0 \left[ \left( \frac{\lambda}{\lambda_0}\right)^{4/3} - 1 \right],
\end{equation}
where
\begin{equation}
\label{eqn:tau0}
    \tau_0 = \frac{1}{c} \left( X \frac{k \lambda_0}{h c}\right)^{4/3} \left[ \left( \frac{G \mbh L_{\textrm{Edd}}}{8 \pi \sigma c^2} \right) (3 + \kappa) \frac{\lambda_{\rm{Edd}}}{\eta} \right]^{1/3},
\end{equation}
where $k$ is Boltzmann's constant and $X$ is a correction factor of order unity, which accounts for the fact that various radii contribute to a single observed time delay at a given wavelength. 

Taking $\kappa = 1$, using the dimensionless accretion rate, $\dot{\mathscr{M}} = 42.9$, in place of $\frac{\lambda_{\rm{Edd}}}{\eta}$, and a mass of $\log \left(\mbh/\msun \right) =7.0$, gives a delay spectrum amplitude of $\tau_0 \approx 0.23 X^{4/3}$. We measured an amplitude of $\tau_0 \approx 2.82$ from the CCF analysis and $\tau_0 \approx 2.32$ from \pyroa, which would require a value of $X = 6.5$ and $X = 5.6$ respectively. The correct value of $X$ is debatable. \citet{Fausnaugh2016} calculated $X=2.49$ from the flux-weighted mean radius, but suggested $X=3.37$ would also be reasonable. Higher values of $X = 4.97$ are suggested by the emission-weighed mean radius rather than the flux-weighted radius \citep[e.g.][]{Edelson2017}. The values of $X$ required to reproduce the observe delay spectrum here are all larger than these theoretical estimates, which suggests the disc size is larger than the theory predicts. Many previous disc reverberation mapping studies have found disc sizes on the order of $\sim$ 2-3 times larger than expected \citep[e.g. ][]{Cackett2020, Edelson2017, Fausnaugh2016}, similar to what we find here. However, this remains an open problem where the magnitude of the overestimation depends on the chosen value of $X$. Equations (\ref{eqn:TempProfile}, \ref{eqn:deltatau}, \ref{eqn:tau0}) are also assuming a thin disc which may not be the case at such high accretion rates, so our inference about the size of the disc is not conclusive.

One solution to the disc size problem was proposed by \citet{Kammoun2021}, where a model delay spectrum was generated from general relativistic ray tracing simulations and successfully reproduces the delay spectra of numerous AGN \citep{Kammoun2021b, Kara2021}, given a corona height and Eddington ratio. We used a mass of $\log \left(\mbh/\msun \right) = 7.0$ and X-ray luminosity of $L_{X,2-10} = 1.67\times10^{43}$ erg s$^{-1}$ as described in Section~\ref{sec:XSpec}. Using {\sc emcee} \citep{Foreman-Mackey2013} we fit the model to the delays obtained with \pyroa\ Model~III and the CCF Model~II delays together, using 15000 samples, 32 walkers and a burn-in of 10000. We used the spin = 1 case, as a higher spin is likely based on the SED modelling. We also include a noise model to scale the errors of the delays, similar to before.

\begin{figure}
	\includegraphics[width=\columnwidth]{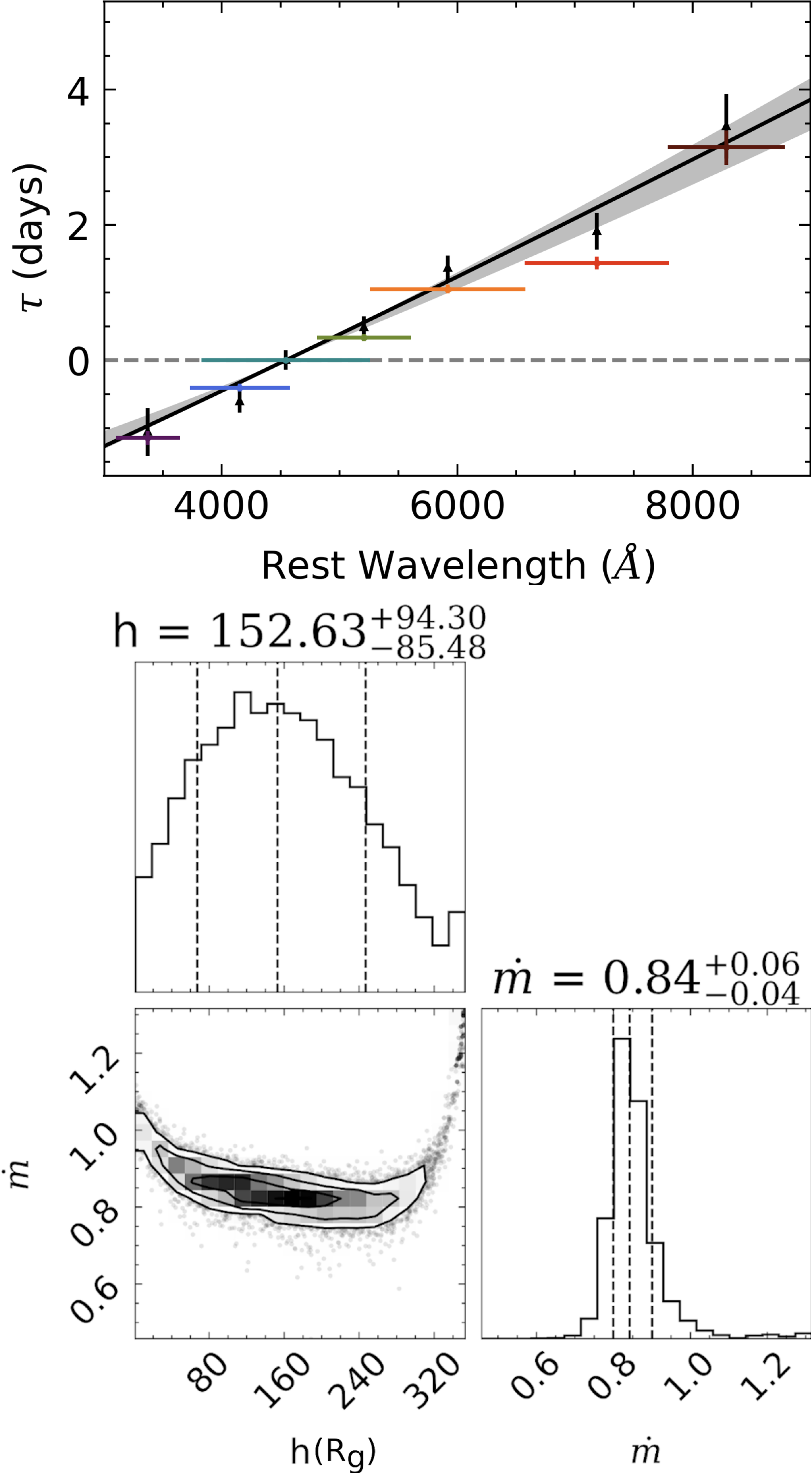}
    \caption{Fit of the \citet{Kammoun2021} model to the combined \pyroa\ Model~III and CCF Model~II delays. {\it Top:} Plot of the delay spectrum with the best fit model. Note, the shape largely does not depend on the specific values of the parameters.  {\it Bottom:} Corner plot of the posterior probabilities of the corona height in gravitational radii and Eddington ratio for a fixed mass of $\log \left(\mbh/\msun \right) = 7.0$.}
    \label{fig:K2020} 
\end{figure}

The results of this fit are shown in Fig.~\ref{fig:K2020}, alongside a corner plot for the posterior probability distributions for the coronal height and Eddington ratio. We find very large coronal heights of $> 100$ R\textsubscript{g}, which are unlikely as estimates from X-ray reverberation experiments suggest heights of $< 10$ R\textsubscript{g} \citep[e.g.][]{Emmanoulopoulos2014, Caballero-Garcia2018}. We show the corner plots in Fig.~\ref{fig:K2020} to demonstrate the degeneracy between this height and the accretion rate where there is a negative correlation towards low heights where the model becomes physical - to fit large delays either the height can be larger and thus the light has further to travel or the accretion rate is larger and thus a larger disc. Therefore, the accretion rate presented in Fig.~\ref{fig:K2020} is likely underestimated as the coronal height is overestimated. To better use this model in the future, the degeneracy needs to be resolved e.g. by measuring the corona height by another method. Incidentally, the corona height of the best fit SED models from \citet{Kubota2018} was $\sim$ 9 R\textsubscript{g}.

In Section~\ref{sec:DelaySpec} we discussed how the theoretical asymmetric response function can cause smaller delays to be measured if not properly modelled. The {\sc CREAM} analysis in Section~\ref{sec:cream} models this precisely assuming a thin accretion disc following a temperature profile of $T(R) \propto R^{-3/4}$. The CREAM fit found a $\log(\mbh\dot{M}/\msun^2\,{\rm yr^{-1}}) = 8.20^{+0.05}_{-0.06}$. From our mass estimate of $\log \left(\mbh/\msun \right) =7.0$, gives an accretion rate of $\dot{M} \approx 15~{\rm M}_{\odot}$ yr\textsuperscript{-1}. This is high, corresponding to $\dot{\mathscr{M}} \approx 27$ which is consistent with the other accretion rate estimates in this work.

To compare the CREAM results with \pyroa\ and CCF, the latter two finding smaller delays, we measured the CREAM delay spectrum amplitude in Section~\ref{sec:DelaySpec}, and is given in Table~\ref{tab:DelaySpecFits} (lags are shown in reference to $g'$ band, as oppose in Fig.~\ref{fig:creamfit} where are shown in respect to the driving lightcurve produced by the lamppost). As expected we find a larger amplitude of $\tau_0 \approx 3.15$ which is significantly larger, $\sim 10 \%$ larger than the CCF results and $\sim 50 \%$ larger than \pyroa\ results.

\subsection{Diffuse Continuum from the BLR}
\label{sec:diffuseBLR}
The \pyroa\ Model~I and CCF I delay for the $u'$ band shows a larger delay than expected for accretion disc reverberation, in particular for the latter, showing a delay larger than the $B$ band. This was only observed when not accounting for a slow varying component and subsequently disappeared after removing the slow variations. This is a common observation \citep[e.g][]{Edelson2019} that has been attributed to ``diffuse continuum'' emission from the BLR, originating at a larger radius and mixing with the disc continuum emission. As this originates from a larger radius, this causes an excess delay. This effect is noticeable in the $u'$ band due to the Balmer jump (3646 \AA) and can also cause excess delay towards the Paschen jump (8204 \AA) \citep[][]{Lawther2018, Korista2019, Netzer2020}. The $u'$ band excess measured by \pyroa\ is weaker, although it also disappears after removing the slow component. This suggests that by this method is less sensitive to this effect, compared to the CCF.

\begin{figure*}
	\includegraphics[width=16cm]{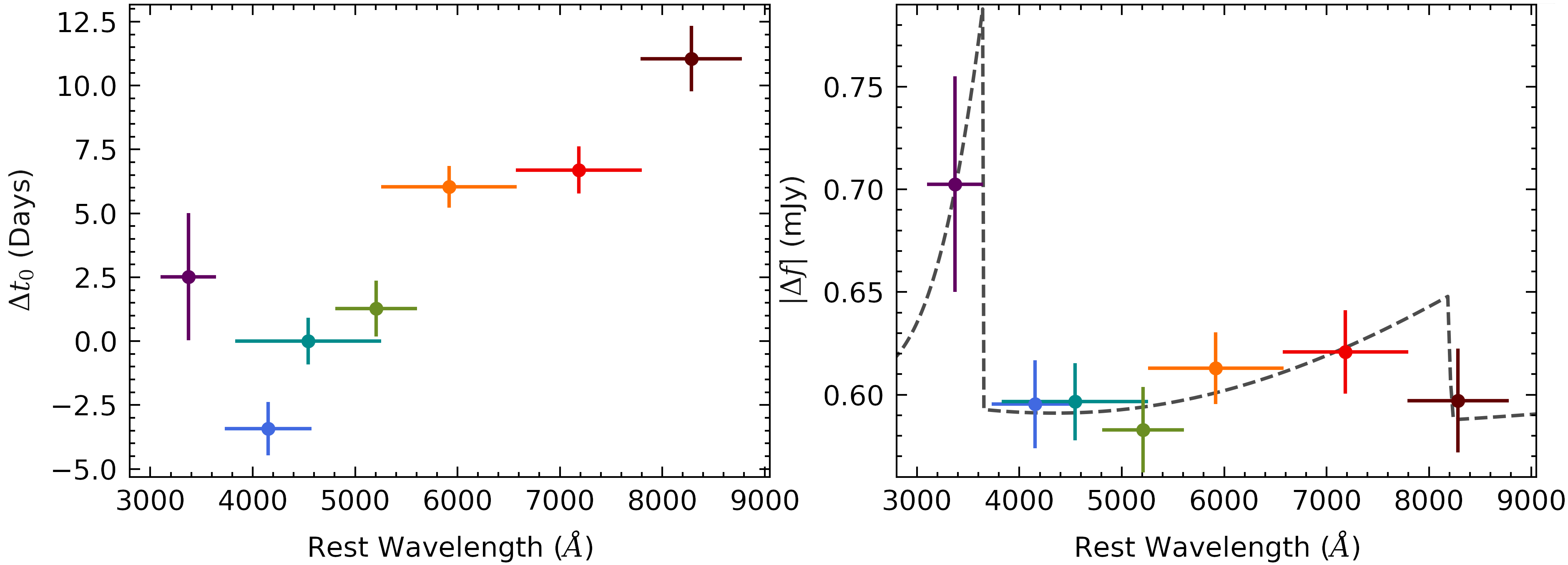}
    \caption{{\it Left:} Difference in the peak time of the slow component for each lightcurve relative to the $g'$ band. The errors in the x-direction indicate the width of the filter. {\it Right:} Amplitude of the slow variations over 100 days for each lightcurve after de-reddening accounting for MW extinction. The grey dashed line shows the shape of hydrogen bound-free emission for reference (this is not a fit to the data).}
    \label{fig:SlowComps} 
\end{figure*}

The slow component was modelled as a parabola given by Eqn.~(\ref{eqn:SlowComp}), where the peak time and change in flux over 100 days are fitted parameters. The left panel of Fig.~\ref{fig:SlowComps} shows the difference in time between the peaks and the $g'$ band. This delay spectrum clearly shows the $u'$ band excess as well as much larger delays than the fast variations shown in Fig.~\ref{fig:DelaySpec}. Larger delays are expected from the BLR as it is at a larger radii, however it may also be due to emission from the accretion disc where the variations are highly smoothed. This can explain the longer delays as the expected delay distribution's for the disc reprocessing model are asymmetric towards larger delays \citep{Starkey2016}, where the driving lightcurve is highly smoothed. Therefore by fitting for only the slow variations, we are measuring delays in the wings of these delay distribution's and therefore appear larger. \citet{Cackett2021} measured delays as a function of Fourier frequency and found that delays are larger at lower frequencies. This is consistent with the larger lags observed here with the low frequency component. Future analysis of these lightcurves, such as a power spectrum analysis could further constrain this component.

The right panel of Fig.~\ref{fig:SlowComps} shows change in flux of the slow component on a timescale of 100 days. The $u'$ band shows a large variation, which quickly drops as wavelength increases before increasing slowly to a peak at $\sim 7300$ \AA. This shape is consistent with the expected spectrum of hydrogen bound-free emission which is shown in the figure. This was generated using the photo-ionisation code CLOUDY version 17.0 \citep{Ferland2017}, using a test suite script for an H~I emission spectrum which provides a reference for the spectral shape. This increased brightness of the $u'$ band can also be easily seen in Fig.~\ref{fig:SEDNew}, where the Swift $U$ band is considerably brighter than the LCO $u'$ band. The Swift data here includes emission from the BLR whereas the LCO data has had this slow component removed (as the sparsity prevents the kind of decomposition that the LCO lightcurves enabled), providing further evidence that the slow variations are originating from the BLR.  To investigate more physical properties of the BLR from this data, more robust modelling is required \citep[e.g ][]{Lawther2018, Korista2019}. All this evidence suggests that these slow variations are driven by diffuse continuum emission from the BLR.

\section{Conclusions}
We have analysed optical continuum lightcurves from the LCO/Swift, X-ray spectra from Swift and optical spectra from Calar Alto to investigate the accretion flow through reverberation mapping of of \pg. Our main findings are:
\begin{enumerate}
    \item From the H$\beta$ lag we measure a black hole mass of $\log \left(\mbh/\msun \right)_{\rm BH} = 7.0$ using the $V$ band continuum. These are consistent but on the lower end of SE mass measurements in the literature.
    
    \item In the continuum lightcurves we detect two components acting on different timescales. The high frequency components are consistent with disc reverberations whereas slow variations are due to diffuse continuum emission from the BLR. By modelling this smooth component we find a delay spectrum with a large Balmer excess and a flux spectrum consistent with that of bound-free hydrogen emission. As this component is very blurred, the emission is likely highly diffuse.
    
    \item We find the disc delay spectrum to be consistent with both a thin disc temperature profile of $T(R) \propto R^{-3/4}$ and the slim disc temperature profile of $T(R) \propto R^{-1/2}$ although the latter less so. The flux spectrum after decomposition is flatter than expected for a thin disc with power $\alpha \sim 0$, possible due to dust reddening.

    \item From modelling the SED, we find an accretion disc that is above the Eddington limit with $\lambda_{\rm{Edd}} =  3.26^{+0.20}_{-0.17}$ for a slim disc and  $\lambda_{\rm{Edd}} = 15.9^{+7.4}_{-5.7}$ for a thin disc, with the slim disc providing a slightly better fit in this case. For both we find a black hole spin close to 1 and the presence of dust, reddening the spectrum, although the thin disc requires more dust to produce a flat spectrum.
    
    \item We tentatively find the disc size to be larger than expected for a thin disc, consistent with other reverberation mapping experiments, however due to many assumptions this is not a conclusive result. 

\end{enumerate}
\pg \ appears to be accreting at a super-Eddington luminosity. Our results do not point to a clear geometry, with both a thin and slim disc being consistent with our data. The slim disc scenario provides a slightly better fit to the SED, but a slightly worse fit to the delay spectrum. From theory, the slim disc scenario is favoured as the high radiation pressure is expected to increase the disc thickness. At the accretion rate measured, the temperature structure may not have fully reached $T(R) \propto R^{-1/2}$, explaining why both models appear to fit the data. Further RM experiments at varying accretion rates are required to better understand this, in particular if objects with higher accretion rates show clearer evidence of a slim/thick disc.


\section*{Acknowledgements}

KH and JVHS acknowledge support from STFC grant ST/R000824/1. CH acknowledges support from the National Science Foundation of China (12122305). PD acknowledges support from NSFC grant 12022301, 11991051, and 11991054, and from National Key R\&D Program of China (grants 2021YFA1600404). LCH was supported by the National Science Foundation of China (11721303, 11991052, 12011540375, 12233001) and the China Manned Space Project (CMS-CSST-2021-A04, CMS-CSST-2021-A06). We thank V. Wild for pointing us towards the Galaxy SED atlas. This work makes use of observations from the Las Cumbres Observatory global telescope network and observations collected at the Centro Astronómico Hispanoen Andalucía (CAHA) at Calar Alto, operated jointly by the Andalusian Universities and the Instituto de Astrofísica de Andalucía (CSIC). We acknowledge the use of public data from the {\it Swift} data archive.
This research made extensive use of {\sc astropy}, a community-developed core Python package for Astronomy \citep{Astropy-Collaboration:2013aa}, {\sc matplotlib} \citep{Hunter:2007aa} and {\sc corner} to visualise MCMC posterior distributions \citep{corner2016}.                                                                                       

\section*{Data Availability}

The raw data can be downloaded from the LCO archive \url{http://archive.lco.global} and the {\it Swift} Archive at \url{https://www.swift.ac.uk/}. The processed lightcurves are available in electronic format via {\sc zenodo } DOI:10.5281/zenodo.7906909.



\bibliographystyle{mnras}
\bibliography{References} 




\appendix

\section{Additional \pyroa\ Fits}

\begin{figure*}
\hspace*{-0.8cm}                                                           
    \centering
	\includegraphics[width=19cm]{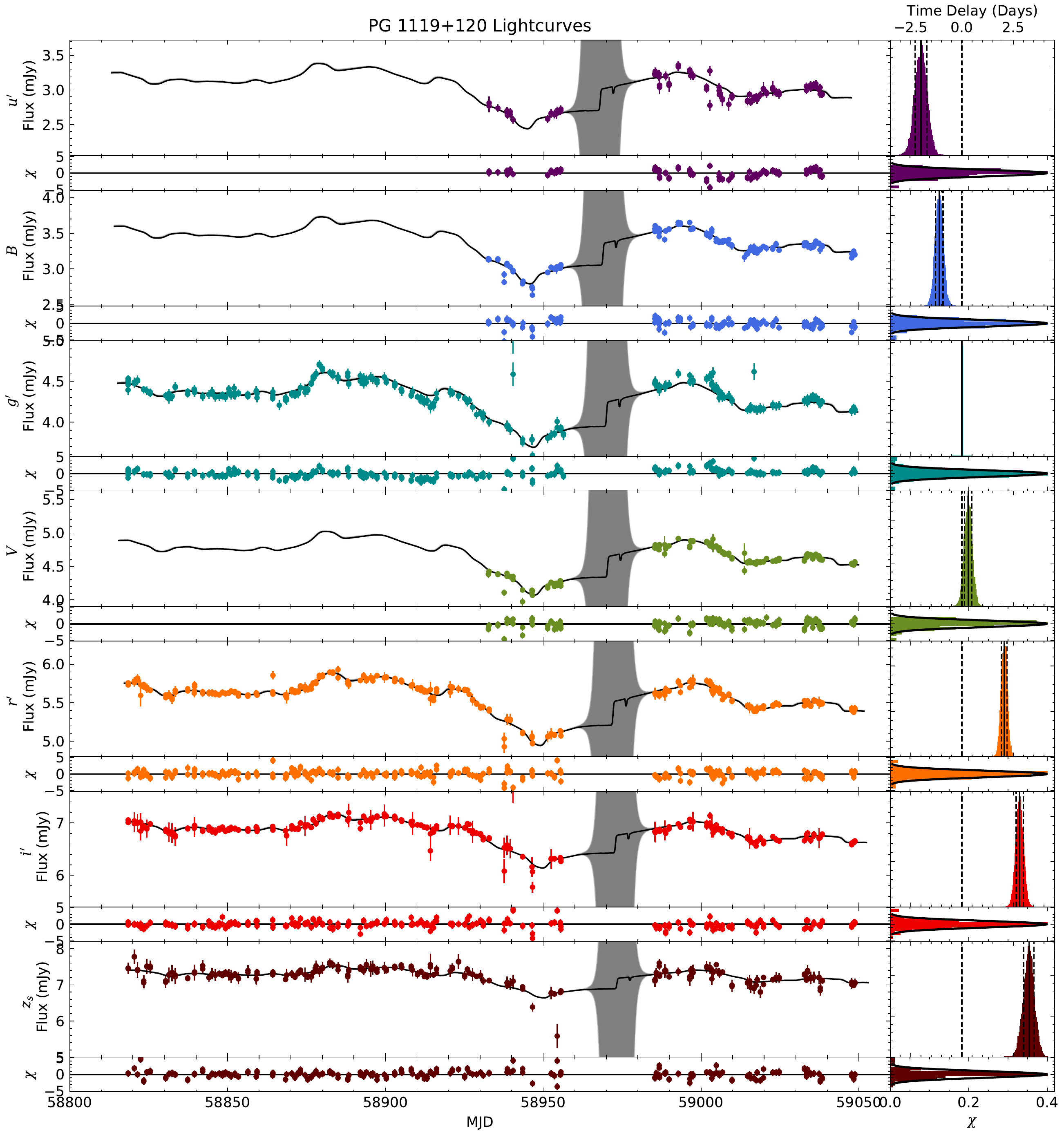}
    \caption{Same as Fig.~\ref{fig:PGFit2} but for Model~I fitted to the first year of data. }
    \label{fig:Year1}
\end{figure*}

\begin{figure*}
\hspace*{-0.8cm}                                                           
    \centering
	\includegraphics[width=19cm]{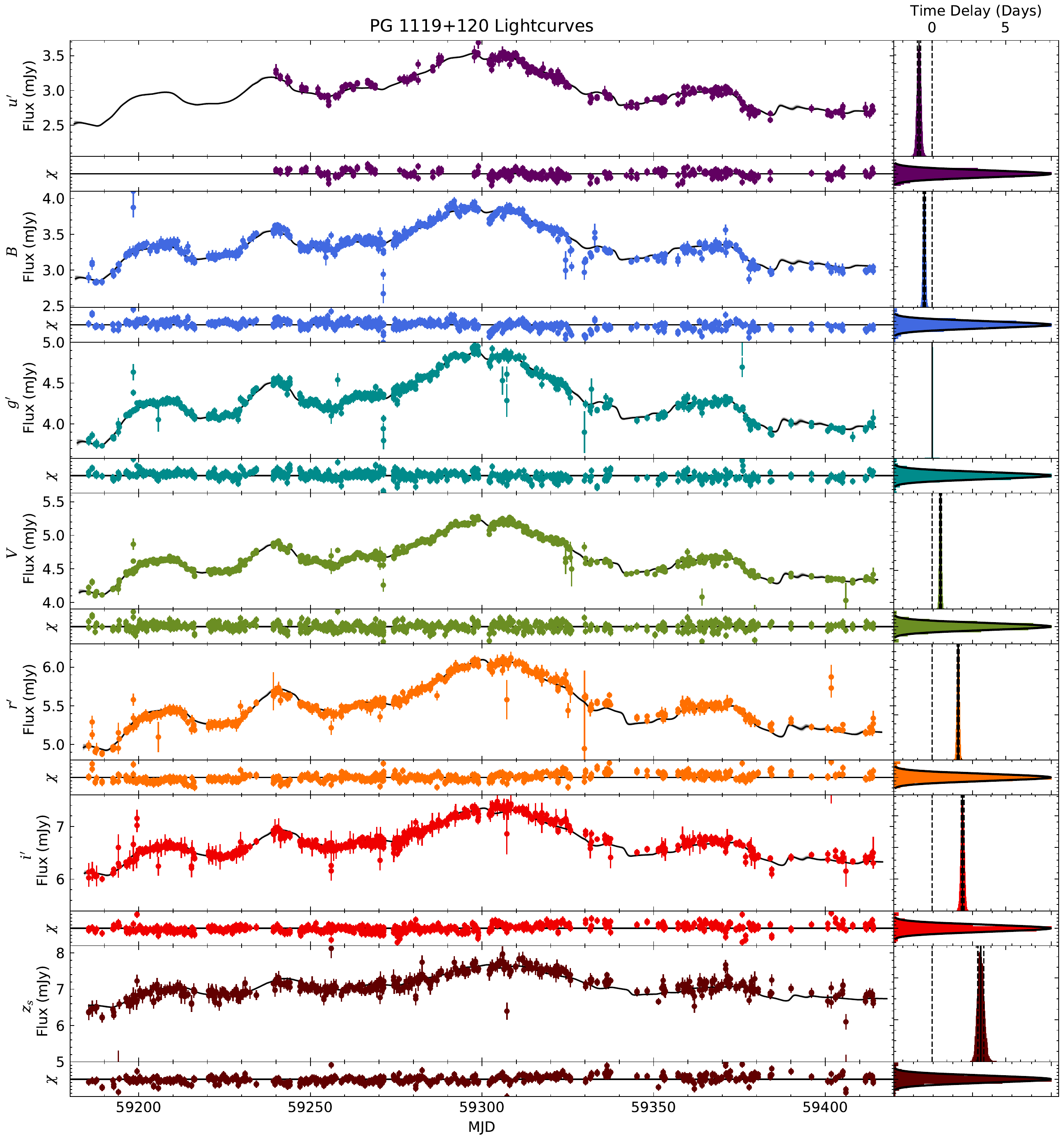}
    \caption{Same as Fig.~\ref{fig:PGFit2} but for Model~I. }
    \label{fig:PGFitModelI}
\end{figure*}
\begin{figure*}
\hspace*{-0.8cm}                                                           
    \centering
	\includegraphics[width=19cm]{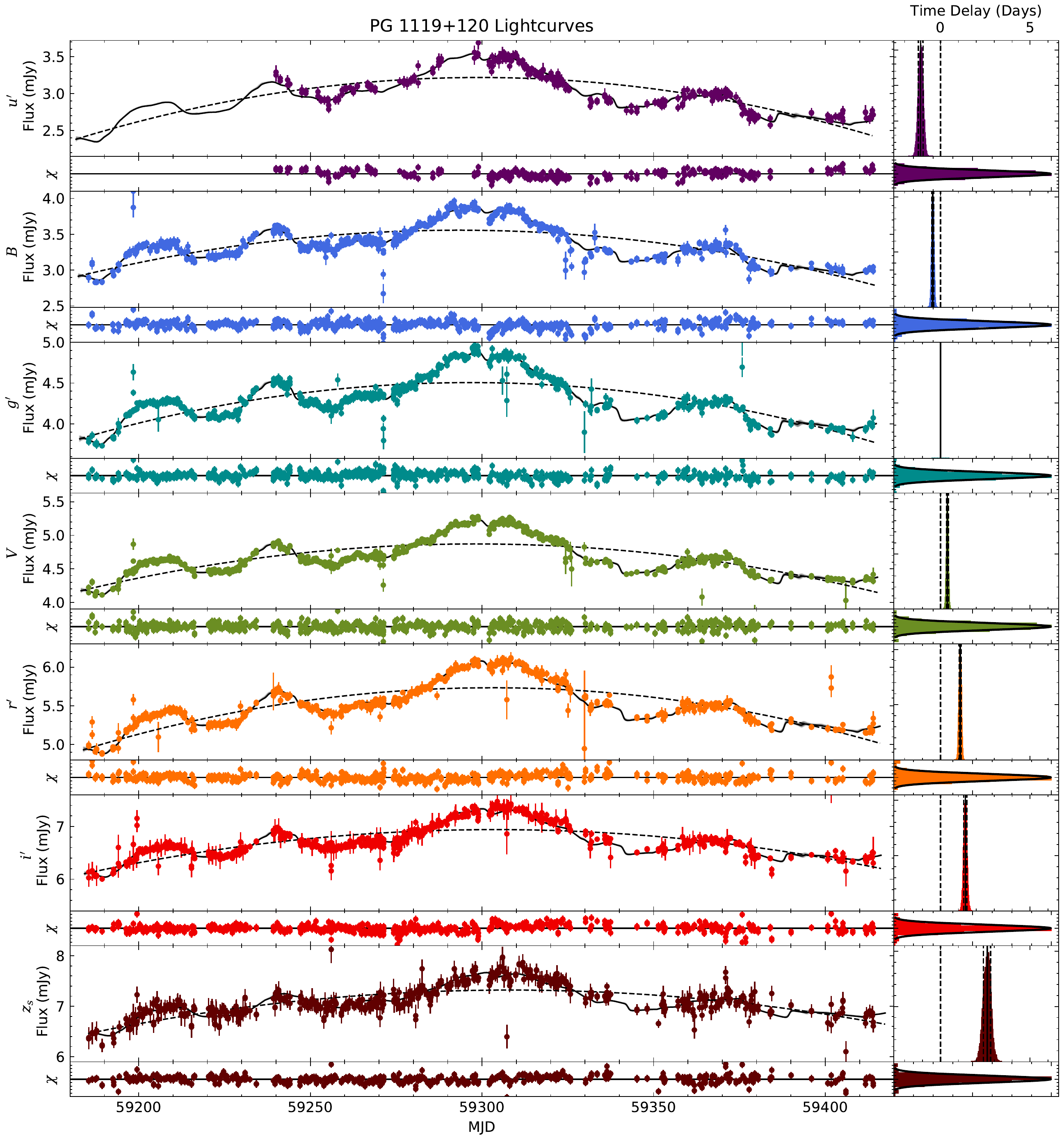}
    \caption{Same as Fig.~\ref{fig:PGFit2} but for Model~II. }
    \label{fig:PGFitModelII}
\end{figure*}

\section{Disc Model Corner Plots}

\begin{figure*}
\hspace*{-0.8cm}                                                           
    \centering
	\includegraphics[width=19cm]{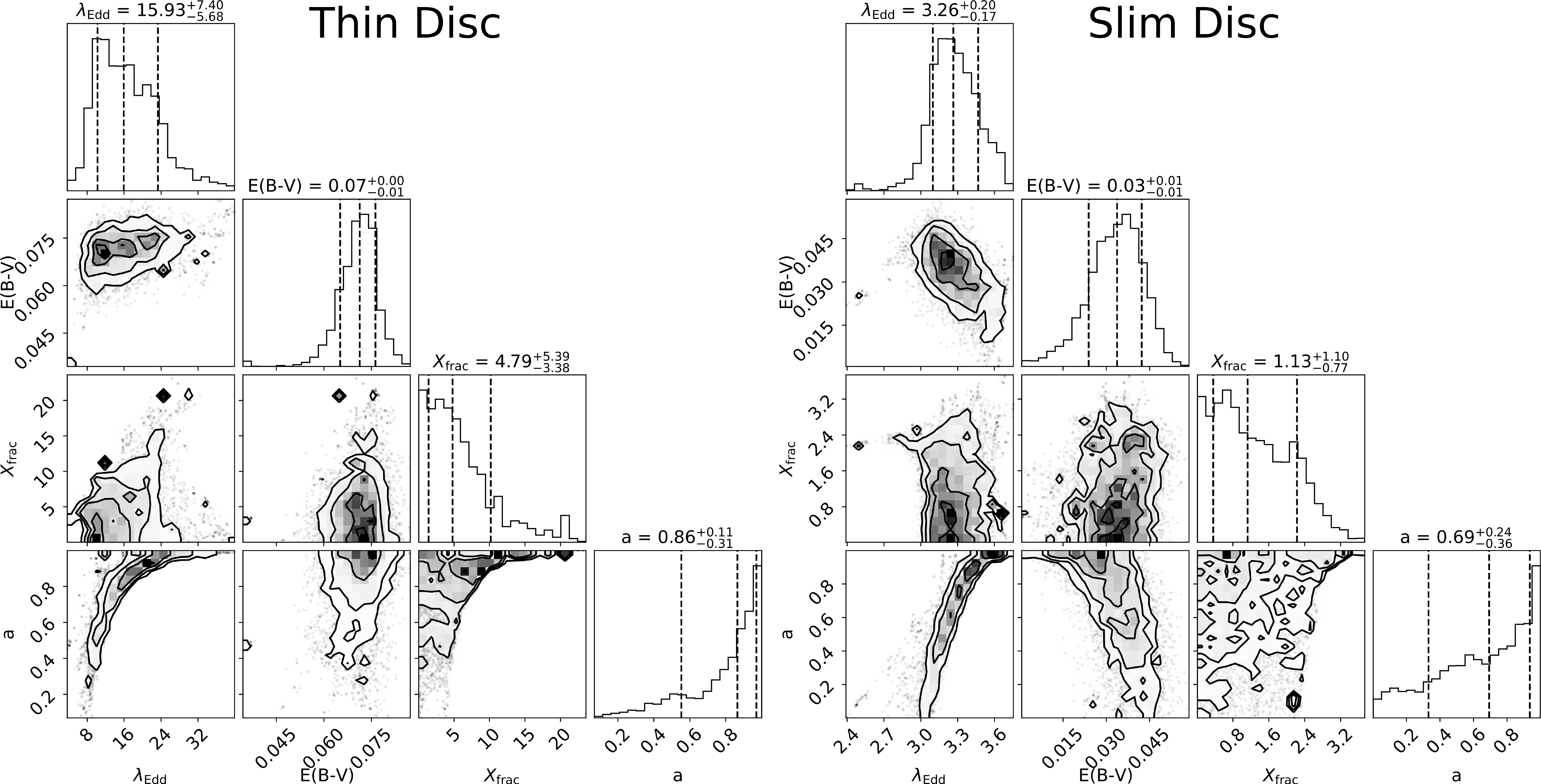}
    \caption{Posterior probability distributions for the thin disc model (right) and the slim disc model (left). A plot of these models is shown in Fig.~\ref{fig:SEDNew}.}
    \label{fig:DiscPost}
\end{figure*}

\section{Lightcurve Model Parameters}

\begin{table*}
\centering
  \caption{Full list of best fit parameters for each \pyroa\ Model. The corresponding figures are referenced in the table}
  \label{tab:PyROAParameters}
    \def\arraystretch{1.2}
    \setlength{\tabcolsep}{10pt}
    \begin{threeparttable}
  \begin{tabular}{ccccccc}
  
    \hline
    Filter, $i$ & $A_i$ & $B_i$ & $\tau_i$ & $s_i$ & $\tau_{\rm{rms}, i}$ \\
     & (mJy) & (mJy) & (Days) & (mJy) & (Days) \\

    (1) & (2) & (3)& (4)& (5)& (6)\\
    \hline
    \multicolumn{6}{c}{Year 1: Model I (Fig. \ref{fig:Year1}), $\Delta = 1.650^{+0.070}_{-0.071}$.}  \\
    \hline
$u'$&$0.213^{+0.004}_{-0.004}$ &$3.046^{+0.005}_{-0.005}$ &$-1.99^{+0.281}_{-0.290}$ &$0.043^{+0.004}_{-0.004}$ &-\\
$B$&$0.212^{+0.003}_{-0.003}$ &$3.393^{+0.003}_{-0.003}$ &$-1.105^{+0.183}_{-0.183}$ &$0.010^{+0.006}_{-0.006}$ &-\\
$g'$&$0.207^{+0.002}_{-0.003}$ &$4.278^{+0.002}_{-0.002}$ &$0.000$ &$0.038^{+0.002}_{-0.002}$ &-\\
$V$&$0.215^{+0.002}_{-0.002}$ &$4.682^{+0.003}_{-0.003}$ &$0.312^{+0.177}_{-0.180}$ &$0.006^{+0.004}_{-0.004}$ &-\\
$r'$&$0.212^{+0.002}_{-0.002}$ &$5.549^{+0.001}_{-0.001}$ &$2.064^{+0.131}_{-0.142}$ &$0.009^{+0.002}_{-0.002}$ &-\\
$i'$&$0.231^{+0.003}_{-0.003}$ &$6.796^{+0.002}_{-0.002}$ &$2.816^{+0.175}_{-0.173}$ &$0.009^{+0.004}_{-0.004}$ &-\\
$z_s$&$0.202^{+0.005}_{-0.005}$ &$7.213^{+0.004}_{-0.005}$ &$3.268^{+0.242}_{-0.267}$ &$0.010^{+0.009}_{-0.007}$ &-\\

    \hline
    \multicolumn{6}{c}{Year 2: Model I (Fig. \ref{fig:PGFitModelI}), $\Delta = 1.184^{+0.037}_{-0.038}$.}  \\
    \hline

$u'$&$0.248^{+0.002}_{-0.002}$ &$2.975^{+0.002}_{-0.002}$ &$-0.889^{+0.111}_{-0.118}$ &$0.025^{+0.003}_{-0.003}$ &-\\
$B$&$0.247^{+0.001}_{-0.001}$ &$3.331^{+0.002}_{-0.001}$ &$-0.533^{+0.065}_{-0.068}$ &$0.029^{+0.002}_{-0.002}$ &-\\
$g'$&$0.274^{+0.001}_{-0.001}$ &$4.268^{+0.001}_{-0.001}$ &$0.000$ &$0.015^{+0.001}_{-0.001}$ &-\\
$V$&$0.266^{+0.001}_{-0.001}$ &$4.635^{+0.001}_{-0.001}$ &$0.566^{+0.052}_{-0.051}$ &$0.007^{+0.002}_{-0.003}$ &-\\
$r'$&$0.279^{+0.001}_{-0.001}$ &$5.472^{+0.001}_{-0.001}$ &$1.770^{+0.063}_{-0.066}$ &$0.033^{+0.001}_{-0.001}$ &-\\
$i'$&$0.303^{+0.002}_{-0.002}$ &$6.668^{+0.002}_{-0.002}$ &$2.062^{+0.099}_{-0.091}$ &$0.046^{+0.002}_{-0.002}$ &-\\
$z_s$&$0.279^{+0.004}_{-0.004}$ &$7.051^{+0.004}_{-0.004}$ &$3.300^{+0.206}_{-0.192}$ &$0.076^{+0.004}_{-0.005}$ &-\\

    \hline
    \multicolumn{6}{c}{Year 2: Model II (Fig. \ref{fig:PGFitModelII}), $\Delta = 1.078^{+0.030}_{-0.032}$.}  \\
    \hline

$u'$&$0.148^{+0.002}_{-0.002}$ &$3.021^{+0.002}_{-0.002}$ &$-1.103^{+0.127}_{-0.132}$ &$0.036^{+0.003}_{-0.003}$ &-\\
$B$&$0.153^{+0.001}_{-0.001}$ &$3.327^{+0.001}_{-0.001}$ &$-0.431^{+0.061}_{-0.059}$ &$0.021^{+0.002}_{-0.002}$ &-\\
$g'$&$0.171^{+0.001}_{-0.001}$ &$4.274^{+0.001}_{-0.001}$ &$0.000$ &$0.012^{+0.002}_{-0.002}$ &-\\
$V$&$0.165^{+0.001}_{-0.001}$ &$4.641^{+0.001}_{-0.001}$ &$0.386^{+0.050}_{-0.052}$ &$0.004^{+0.003}_{-0.002}$ &-\\
$r'$&$0.160^{+0.001}_{-0.001}$ &$5.487^{+0.001}_{-0.001}$ &$1.100^{+0.063}_{-0.062}$ &$0.017^{+0.001}_{-0.001}$ &-\\
$i'$&$0.182^{+0.002}_{-0.002}$ &$6.686^{+0.002}_{-0.002}$ &$1.387^{+0.085}_{-0.088}$ &$0.034^{+0.002}_{-0.002}$ &-\\
$z_s$&$0.160^{+0.004}_{-0.004}$ &$7.069^{+0.004}_{-0.004}$ &$2.609^{+0.190}_{-0.212}$ &$0.060^{+0.004}_{-0.004}$ &-\\
    \hline
    \multicolumn{6}{c}{Year 2: Model III (Fig. \ref{fig:PGFit2}), $\Delta = 0.991^{+0.033}_{-0.032}$.}  \\
    \hline
$u'$&$0.154^{+0.002}_{-0.002}$ &$3.041^{+0.002}_{-0.002}$ &$-1.206^{+0.113}_{-0.105}$  &$0.033^{+0.002}_{-0.002}$&$0.000$ \\
$B$&$0.155^{+0.001}_{-0.001}$ &$3.349^{+0.001}_{-0.001}$ &$-0.423^{+0.06}_{-0.061}$  &$0.017^{+0.002}_{-0.002}$ &$0.027^{+0.04}_{-0.02}$\\
$g'$&$0.169^{+0.001}_{-0.001}$ &$4.291^{+0.001}_{-0.001}$ &$0.000$ &$0.010^{+0.002}_{-0.002}$ &$0.208^{+0.115}_{-0.106}$ \\
$V$&$0.166^{+0.001}_{-0.001}$ &$4.665^{+0.001}_{-0.001}$ &$0.352^{+0.054}_{-0.053}$ &$0.007^{+0.002}_{-0.003}$ &$0.770^{+0.094}_{-0.088}$ \\
$r'$&$0.157^{+0.001}_{-0.001}$ &$5.510^{+0.001}_{-0.001}$ &$1.103^{+0.068}_{-0.066}$  &$0.017^{+0.001}_{-0.001}$ &$1.304^{+0.123}_{-0.107}$\\
$i'$&$0.180^{+0.002}_{-0.002}$ &$6.705^{+0.002}_{-0.002}$ &$1.508^{+0.103}_{-0.097}$ &$0.032^{+0.002}_{-0.002}$ &$1.919^{+0.181}_{-0.173}$ \\
$z_s$&$0.161^{+0.004}_{-0.004}$ &$7.089^{+0.004}_{-0.003}$ &$3.309^{+0.261}_{-0.274}$  &$0.053^{+0.006}_{-0.008}$&$2.402^{+0.368}_{-0.363}$ \\
    \hline

  \end{tabular}
\begin{tablenotes}
    \item[] (1): Lightcurve filter. (2): Lightcurve RMS. (3): Lightcurve mean. (4): Time delay. (5): Noise model where flux variance is increased by $s_i^2$. (6): Log-Gaussian delay distribution width (see equations (\ref{eqn:DelayDist}, \ref{eqn:DelayDistConv})).
    \item[] The values of the ROA window width, $\Delta$, are shown in the table headings.
  \end{tablenotes}
  \end{threeparttable}
 \end{table*}

\bsp	
\label{lastpage}
\end{document}